\documentclass[journal=jacsat,manuscript=article]{achemso}

\usepackage{chemformula} 
\usepackage[T1]{fontenc} 

\usepackage{amsmath}
\usepackage{amssymb}
\usepackage{float}
\usepackage{lineno}
\usepackage{wrapfig}
\usepackage{subcaption}
\usepackage{setspace}
\usepackage{epsfig}
\usepackage{amsthm}
\usepackage{lscape,graphicx}
\usepackage{array}
\usepackage{caption}
\usepackage{setspace}
\usepackage{fullpage}
\usepackage{centernot}
\usepackage{bbm}
\usepackage{chemarr}
\usepackage{changes}
\usepackage{tcolorbox}
\newcommand{\real}{\mathbb{R}}

\newcommand{\utwi}[1]{\mbox{\boldmath $ #1$}}

\newcommand{\bx}{{\utwi{x}}}

\newcommand{\bH}{{\utwi{H}}}

\newbool{comments}
\booltrue{comments}

\author{Farid Manuchehrfar}
\altaffiliation{These authors contributed equally}

\author{Huiyu Li}
\altaffiliation{These authors contributed equally}

\author{Wei Tian}
\altaffiliation{These authors contributed equally}

\author{Ao Ma}
\email{aoma@uic.edu}

\author{Jie Liang}
\email{jliang@uic.edu}

\affiliation{Center for Bioinformatics and Quantiative Biology and Department of Bioengneering, University of Illinois at Chicago, Chicago, IL 60607.}

\title[Exact Topology of an Activated Process]
  {Exact Topology of Dynamic Probability Surface of an Activated Process by Persistent Homology}


\begin{document}




\begin{abstract}
To gain insight into reaction mechanism of activated processes, we introduce an exact approach for quantifying the topology of high-dimensional probability surfaces of the underlying dynamic processes. Instead of Morse indexes, we study the homology groups of a sequence of superlevel sets of the probability surface over high-dimensional configuration spaces using persistent homology.  For alanine-dipeptide isomerization, a prototype of activated processes, we identify locations of probability peaks and connecting-ridges, along with measures of their global prominence. Instead of a saddle-point, the transition state ensemble (TSE) of conformations are at the most prominent probability peak after reactants/products, when proper reaction coordinates are included. Intuition-based models, even those exhibiting a double-well, fail to capture the dynamics of the activated process. Peak occurrence, prominence, and locations can be distorted upon subspace projection. While principal component analysis account for conformational variance, it inflates the complexity of the surface topology and destroy dynamic properties of the topological features. In contrast, TSE emerges naturally as the most prominent peak beyond the reactant/product basins, when projected to a subspace of minimum dimension containing the reaction coordinates. Our approach is general and can be applied to investigate the topology of high-dimensional probability surfaces of other activated process.

\end{abstract}

\section{Introduction}

Activated processes are ubiquitous in molecular systems, ranging from chemical reactions of small molecules to  dynamic conformational changes  and enzymatic reactions of proteins.  In proteins, all functionally important processes are activated processes, 
which provide well-defined rates 
essential for proteins to carry out their  roles in the cellular context, as proper timing is required for proper function. 

The prevalent picture describing an activated process is that of a transition between two meta-stable basins on the free energy landscape separated by a barrier, whose height is large compared to thermal energy~\cite{Chandler1978}.  The slow time scale of activation arises from the fact that the molecular system can  rarely  accumulate sufficient energy in the relevant degrees of freedom (DoFs) to surpass the transition barrier.  This simple and 
elegant
picture originates from reaction rate theories, such as the well-known transition state theory and Kramer's theory~\cite{Chandler1978, Kramers1940284, Pechukas1976, Wigner1938, Hanggi1990,berne1988, Pollak2005} developed in studies of the dynamics of chemical reactions of small molecules.  

A key concept in reaction rate theories is that of reaction coordinates: a few special coordinates exists that can fully determine the progress of a reaction process \cite{Bolhuis2002,Du1998, Li_Ma2014}. 
A requirement for reaction coordinates is that they must accurately locate the transition barrier.  Accordingly, the numerous degree of freedoms (DoFs) in a complex molecular system (\textit{e.g.}, a protein molecule, a system of solute and solvent) can be
divided into reaction coordinates and heat bath.  Reaction coordinates play  central roles as they determine both the mechanism and the rate of activation.  For example, to modify the activity of an enzyme, one should modify residues involved in the reaction coordinates of  the enzyme activities~\cite{Schramm2018, Schwartz2009}, as this will modify both the reaction pathway and the barrier height for activation.  In contrast, modifying residues that belong to the heat bath will not alter the enzymatic activity, as the role of the heat bath is to provide energy to the reaction coordinates to cross the activation barrier during rare fluctuations, which is 
largely a non-specific process.

Given such significance, it is important to develop a rigorous and quantitative criterion for determining the correct reaction coordinates.  This task was accomplished with the development
of 
the procedure of \textit{committor test}, which is characterized by the committor value $p_B$~\cite{Bolhuis2002,Du1998,RYTER1987,Onsager1938}: the probability that a dynamic trajectory initiated from a given configuration to reach the product basin before visiting the reactant basin.  By definition, the reactant and product states have committor values of 0 and 1, respectively, whereas the optimal transition state coincides with  $p_B=0.5$.
The committor value $p_B$ therefore provides a rigorous parameterization of the reaction process.  Thus, the intuitive, albeit qualitative, notion of reaction coordinates translates into a rigorous definition of the few coordinates that are sufficient for determining the committor value of any given configuration.  Du \textit{et.\ al} first adopted this rigorous definition of reaction coordinates in the context of protein folding~\cite{Du1998}; Chandler and co-workers established its usage as a standard practice in the general context of activated processes~\cite{Bolhuis2002}.

While this rigorous criterion has been well accepted,  identifying the correct reaction coordinates turns out to be rather difficult, even for systems of modest complexity.  One example is the $C_{7eq} \rightarrow C_{7az}$  isomerization reaction of the alanine dipeptide in vacuum, a prototype for studying biomolecular conformational transitions.  Alanine dipeptide is the smallest molecule that satisfies the criterion that distinguishes complex molecules from small molecules: the non-reaction coordinates in the system constitute a large enough thermal bath to provide the reaction coordinates with adequate energy to cross the activation barrier.  As a result, the $C_{7eq} \rightarrow C_{7az}$ transition displays features of activated dynamics that are unique to complex molecules but absent in small molecules.  It was first found by Bolhuis~\textit{et.\ al}~\cite{Bolhuis2000} that the conventional Ramachandran torsional angles $\phi$ and $\psi$, while sufficient for distinguishing the two stable basins, are inadequate for locating the transition state.  Instead, another torsional angle $\theta_1$ was found to be an essential reaction coordinate––a rather counter-intuitive finding~\cite{Li_Ma2020KineticEnergy, Li_Ma2016_Reaction_Mechanism}.  The counter-intuitive nature of reaction coordinates turned out to be more often the norm than exception in complex systems~\cite{Bolhuis2000,Best2005,Antoniou2011}, posing a formidable challenge, as sans intuition, there appears no guidance in sight.

The challenge in identifying reaction coordinates had motivated  efforts in developing 
rigorous methods for their identification  in complex systems since the early 2000’s~\cite{Bolhuis2002,Du1998, Bolhuis2000,Best2005, Hu2008,Ma2005Automatic}.  Beyond unreliable intuition and  trial-and-error, the first systematic method was that of machine-learning, in which a neural network was used to automatically identify the optimal reaction coordinates from a prepared pool of candidates~\cite{Ma2005Automatic}.  This method was used to successfully identify the key solvent coordinate that controls the isomerization dynamics of an alanine dipeptide in solution, which had defied prior intuition-based trial-and-error efforts. 
The success of this machine learning based approach lead a series of further developments along similar lines~\cite{Li_Ma2014, Best2005, Peters2006, Antoniou2009, Covino2019, Sidky2020,Bonati2019,Wang2019,WANG2020ML, Antoniou2011}.  

However, a major deficiency of machine learning-based methods is that they cannot answer the real question concerning reaction coordinates – why some coordinates are more important for activation than the others?  Instead, these methods can only inform us empirically which coordinates appear to be important based on well-defined criteria.  Recently, Ma and co-workers developed a rigorous theory for mapping out the flow of potential energy through individual coordinates~\cite{Li_Ma2016TPS,Li_Ma2016_Reaction_Mechanism}.  It was found that the reaction coordinates are the coordinates that carry high energy flows during the activation process.  This result suggested an appealing physical picture: energy flows from fast coordinates into slow coordinates during activation, so that adequate energy can accumulate in the slow coordinates,  enabling them to cross the activation barrier on path of these slow coordinates.  This physical picture also suggested that reaction coordinates are preferred channels of energy flows and  are encoded in the protein structure. Through analysis of energy flow, one can obtained a prioritized list of coordinates that likely play most significant roles in the activation process.

The most celebrated concept in reaction rate theories is that of the transition state (TS), which is the dynamic bottleneck of an activated process.  Conventional thoughts are that they are located at a critical point with Mores index of 1 on the high-dimensional potential energy landscape of the molecule.  
If all the reaction coordinates of an activated process are known, the TS will be an index-1 critical point on the free energy surface of the reaction coordinates, the sole direction with negative Hessian pointing along the ideal one-dimensional reaction coordinate valid at the top of the activation barrier.  Based on this picture, the surface of the multi-dimensional probability distribution of reaction coordinates constructed from an ensemble of reactive trajectories should be highly structured, with the important dynamic states (\textit{e.g.} TS) manifesting as critical topological features.

To gain insights into reaction mechanism, a common practice is to study features of the free energy surface.  However, 
all relevant information of an activated process is contained in the reactive trajectories. 
Instead of the free energy surface, one can  construct the \textit{dynamic probability surface} of the transition state.  For certain systems, this can be achieved using the transition path sampling (TPS) method. Once a large ensemble of reactive trajectories
are generated, an ensemble of configurations that the system sampled during the transition process can be harvested.  From this ensemble, one can generate a dynamic probability landscape or transition state surface, which is usually high-dimensional.

The focus of this work is the analysis of the exact topology of the high-dimensional dynamic probability landscape, and the establishment of  its relationship with the transition state of the active process. 
Instead of Morse index, we characterize the high-dimensional transition state surface by its topological structures in homology groups. We analyze topological changes
 in the superlevel set of the probability surface at different probability levels. Using the technique of persistent homology, 
we identify the locations of probability peaks in the high-dimensional configuration space and ridges connecting them, along with measures of their global prominence.

We apply this approach to study 
the active process of
$C_{7eq}\rightarrow C_{7ax}$ isomerization of the alanine dipeptide, a well-characterized model system for studying protein conformational changes~\cite{Bolhuis2000,Li_Ma2016TPS,Li_Ma2016_Reaction_Mechanism}.
After the exact topological structures of the dynamic probability surface are constructed, we identify the location of the ensemble of transition state conformations.
Instead of a saddle point with a Morse index of 1,  the transition state ensemble (TSE) is found to be located at the top of the most prominent peak after those of the reactant and product.
In addition, the dynamically important topological structures are  retained  when  the  surface is projected onto the 2-dimensional  plane of ($\phi-\theta_1$), which are  known to be the reaction coordinates~\cite{Li_Ma2016_Reaction_Mechanism}. 
In contrast, when projected to the intuition-derived ($\phi-\theta$)-plane, the topological features of the  probability surface no longer contain dynamic information on  the transition state.  
Furthermore, we find that PCA dimension reduction distorts surface topologies, such that the transition state ensemble cannot be recovered from PCA-derived topological features:  Instead of simplification, PCA destroys the dynamic properties of topological features of the original transition state surface.

Overall, we introduce a novel approach for quantifying the exact topology of high-dimensional surfaces.  
With this approach, we have characterized the precise topology of the transition state surface of the active process of alanine dipeptide isomerization.  We have also established that the TSE are located at the most prominent probability peak beyond those of the reactant and the product, when the  subspace of projection contain the proper reaction coordinates.  The new language of homology group and the technique of persistent homology on superlevel sets of high-dimensional probability surfaces introduced here 
are general and can be applied to investigations of the  topology of high-dimensional surfaces over configuration space encountered in other problems of activated process.

\section{Theory, Models and Methods}

We briefly discuss how the dynamic probability surface can be constructed for the molecular system of our interests. We then 
discuss the problem of understanding  
118

the topological structures of the probability surface over the relevant configuration space. 
Our focus will be the introduction of the method of homology groups for analyzing the exact topological structures of the dynamic probability surface.  This approach is based on the homological structures of the sequence of the superlevel sets of a probability surface, and differs from previous efforts that are based on critical points, Morse theory, and Euler characteristics.

\subsection{Constructing dynamic probability surface: transition state ensemble and path sampling}
We use the transition path sampling (TPS) method~\cite{Bolhuis2002} to generate a sufficiently large ensemble of
reactive trajectories for the active process we study.  To ensure the transition process is fully covered without bias, the duration of the trajectories is much longer than that of the transition process.
Along each trajectories, we further harvested a number of system configuration with a specific time intervals %
to generate an ensemble of the relevant configurations that the system samples during the transition process. 
From this ensemble, we construct a dynamic probability landscape over a $d$-dimensional subspace, namely,  the joint probability distribution over the $d$-coordinates.

\subsection{Configuration space and probability surface.}
We now discuss the general problem of analyzing probability surface over a configuration space of arbitrary dimension.  We first  introduce a few relevant concepts, and then review current approaches.  This is followed by an exposition of the basics of homology groups and persistent homology.  We focus on developing these concepts in the setting of cubics and superlevel sets of the probability surface.  We will also discuss the key concept  of filtration and its  construction, and how it can be used to uncover the exact high-dimensional topological features. 

\paragraph{Configuration space.} We begin our discussion in a general setting. We use $d$ number of
features that describe the configurations of a molecule.  These can be bond lengths, bond angles, and torsional angles describing the structure of the molecule.  For alanine dipeptide, there are a total of $60$ features that fully describes the configuration of the molecule.  For a molecule of finite size, the configuration space $\mathbb{M} \subset \real^d$ is  compact.  
It is likely that $\mathbb{M}$ lies in a subspace of the Euclidean space  $\real^d$, as there are coupling between different degrees of freedom.  

\paragraph{Probability surface.}
Each configuration $\bx = (x_1,\, x_2, \cdots, x_d)  \in \mathbb{M}$ has a probability $f(\bx) \in [0,\,1]$ associated with it. Namely, we have a function
$
f: \mathbb{M} \rightarrow \real_{[0,\,1]}
$
that assigns the probability value $p(\bx)$ to each configuration $\bx$.

\paragraph{Superlevel sets and sublevel sets.}
For a value $0\le a \le 1$, we can identify all $\bx$ whose probability values 
$f(\bx)\ge a$, which is called the
\textit{superlevel set} $\mathbb{M}_{f\ge a}$:
$$
\mathbb{M}_{f\ge a} \equiv \{ \bx \in \mathbb{M}| f(\bx) \ge a\} = f^{-1}([a,\,1)).
$$
Similarly, we can have the
the 
\textit{sublevel set} $\mathbb{M}_{f\le a}$:
$$
\mathbb{M}_{f\le a} \equiv \{ \bx \in \mathbb{M}| f(\bx) \le a\} = f^{-1}((0,\,a]).
$$

\subsection{Topology of probability surface: Critical points and Morse indices.} 

It is of great importance to understand the topological structures of the probability surface $f(\bx)$.
The approach of analyzing the critical points 
is well-practiced for exploring the topological structures of high-dimensional surfaces.

\paragraph{Critical Points.}
The critical points of $f(\mathbb{M})$ are where all first derivatives of $f$ vanishes:
$$
\frac{\partial f(\bx) }{\partial x_1} =0,\,
\frac{\partial f(\bx) }{\partial x_2} =0,\,
\cdots,\,
\frac{\partial f(\bx) }{\partial x_d} =0.
$$
Critical points are coordinates independent and can be further classified into different types by the secondary derivatives. We can organize the secondary derivatives into a $d\times d$ Hessian matrix $\bH f(\bx)$, whose entries are:
$$
(\bH f)_{i,\,j} = \frac{\partial^2 f}{\partial x_i \partial x_j}.
$$
At non-degenerative critical points, where the Hessian matrices are non-singular, the Hessian will have a mixture of positive and negative eigenvalues.  The number of negative eigenvalues $\eta$ is the \textit{Morse index} of the critical point.

\paragraph{Topology of surface by critical points and Morse theory.}
At a critical point, the topology of sublevel sets  changes. For a critical point $\bx$ with $f(\bx) = a$,  consider the  sublevel sets $\mathbb{M}_{f<a}$ and the sublevel set $ \mathbb{M}_{f<(a+\epsilon)}$ slightly above it by a small amount of $\epsilon>0$.
The topology of $\mathbb{M}_{f<a}$ and $ \mathbb{M}_{f<(a+\epsilon)}$ are different, as one cannot be deformed into another.
However, once an $\eta$-dimensional handle is attached to 
$\mathbb{M}_{f<a}$
at the critical point $\bx$, it has the same homotopy type as 
$\mathbb{M}_{f<(a+\epsilon)}$~\cite{matsumoto2002introduction}, namely, these two can be deformed into one another. In fact, the homotopy class of  $\mathbb{M}$ can be characterized by toplological changes at critical points. The problem of determining the topology of the surface $f(\bx)$ on $\mathbb{M}$ then becomes the problem of determining the critical points of $f(x)$ on $\mathbb{M}$ and their Morse indices.

\paragraph{Euler characteristics.}
A special case of the celebrated Morse inequalities relates the number of critical points of different indices to the Euler characteristics, which is
$$
\chi(\mathbb{M}) = \sum_{k=1}^d (-1)^k m_k,
$$
where $m_k$ is the number of critical points of index $k$.  The Euler characteristics provides information on the number of holes of various dimensions, and can also be written as:
$$
\chi(\mathbb{M}) = \sum_{k=1}^d (-1)^k \beta_k,
$$
where $\beta_k$ is the $k$-th Betti number, which counts the number of $k$-dimensional holes.  For example, in $\real^3$,
$\beta_0$ counts the number of connected components, 
$\beta_1$ counts the number of tunnels, 
and 
$\beta_3$ counts the number of voids. 
Voids and tunnels have been extensively examined in the studies of protein structures and functions~\cite{liang1998a,Liang1998b, Liang_EDELSBRUNNER1998a,Liang_Edelsbrunner1998b,binkowski2003inferring,binkowski2005protein,tseng2006estimation,tseng2009predicting,perez2018mechanism,tian2018castp}.

Persistent homology has also been applied in analysis of structures of chemical compounds and proteins, and for correlation with their biophysical properties as well as data analysis~\cite{xia2014persistent,xia2015persistent,xia2015multidimensional}.

\paragraph{Prior work on characterizing topological surface of molecular system.}
The topology of surfaces over configuration space has been the subject of many investigations~\cite{caiani1997geometry,angelani2005relationship,kastner2008phase,wales2018exploring,cimasoni2019topological} .  For example,
thermodynamic phase transitions have been explored from the viewpoint of topological changes  of  submanifolds  of  the configuration space.  Changes in Euler characteristics have been found to signal the occurrence of phase transitions for certain systems~\cite{angelani2005relationship,kastner2008phase}. 

However, it is difficult in practice  to understand the global surface topology using critical points and Morse indices.  
There are only a few model systems where all critical points are known analytically~\cite{kastner2011phase}.  Numerical computation faces numerous challenges.
First, it is difficult to identify all critical points.
Methods based on Newton-Ralphson and other techniques require initial guesses and do not guarantee identifications of all critical points.  Furthermore, many initial guesses fall into the same large basins of attraction and yield no new information.
Second, as the probability surface can only be constructed approximated from points sampled in the high-dimensional configuration space, the degree of sampling may not be sufficiently detailed to capture the original topology of the configuration space. 
Third, sampling has to be sufficiently detailed  to accurately measure the first and second derivatives along each coordinate direction at all locations.   
Fourth, as derivatives reflect local properties of the probability surface, there can be numerous critical points that may be trivial and of little importance if the probability surface is rugged. It is difficulty to distinguish those reflecting the global features of the surface from those reflecting local dimples  or even those that are due to noise in sampling. While there have been efforts in visualizations of potential surface of molecules and other heuristics~(see~\cite{wales2018exploring} and references within),  to our knowledge, it is not yet possible to characterize all critical points and the homotopy type of the probability or potential surface of a  molecule in three dimensional space.

\subsection{Topology of probability surface: Homology group and persistent homology }
\paragraph{Background and overview.}

Instead of analyzing the critical points, we adopt a different approach. We are interested in global features  such as  the occurrence of different peaks, and how they are connected. 
This can be achieved by examining globally the structures of holes of various dimension and how such structures changes at different sublevel or superlevel sets of the probability landscape. 
We adopt an approach  based on the theory of homology group and persistent
homology~\cite{Edelsbrunner:2230405,Edelsbrunner2002, carlsson2009topology}. Homology group studies holes in topological spaces and is a classic topic in algebraic topology~\cite{munkres2018elements,hatcher2005algebraic}.
Persistent homology computes these holes and measures their scales  at different spatial
resolutions~\cite{Edelsbrunner2002, carlsson2009topology}.
Compared to homotopy, homology groups are more amenable to computation.  Our approach is only feasible due to recent progress in computational topology and topological data
analysis~\cite{Edelsbrunner2002, Cohen-Steiner-stability, carlsson2009topology, Edelsbrunner:2230405,wagner2012efficient}. 

\begin{figure}[h!]
    \begin{center}
     \includegraphics[width=0.99\textwidth]{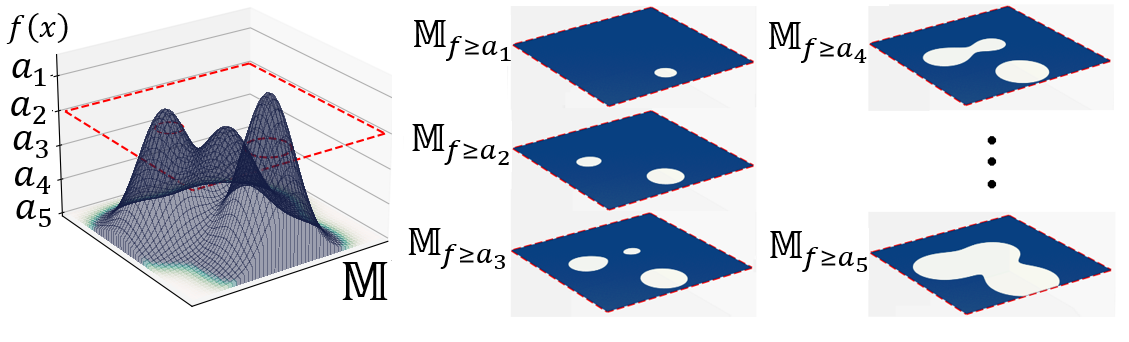}
    \end{center}
   \caption{\sf    Sea levels of the probability landscape $f(\bx)$ on the configuration space $\mathbb{M}$. The superlevel sets $\mathbb{M}_{i}=\mathbb{M}_{f\ge a_i} =   f^{-1}(\ge a_i)$  at different sea level (white regions) can have different topology, with different number of components shown in white.}    \label{fig:seaLevels}
\end{figure}

To provide  an intuitive picture illustrating this approach, we envision a sea level on top of the probability landscape over the configuration space~(Fig~\ref{fig:seaLevels}). We are interested  in how mountain peaks emerge from the sea when the sea level is lowered gradually, and how independent mountain peaks become connected by land-ridges when the sea level is lowered further. 
These are related to $0$-dimensional holes, which are connected components.

\begin{figure}[h!]
    \centering
    \includegraphics[scale=0.37]{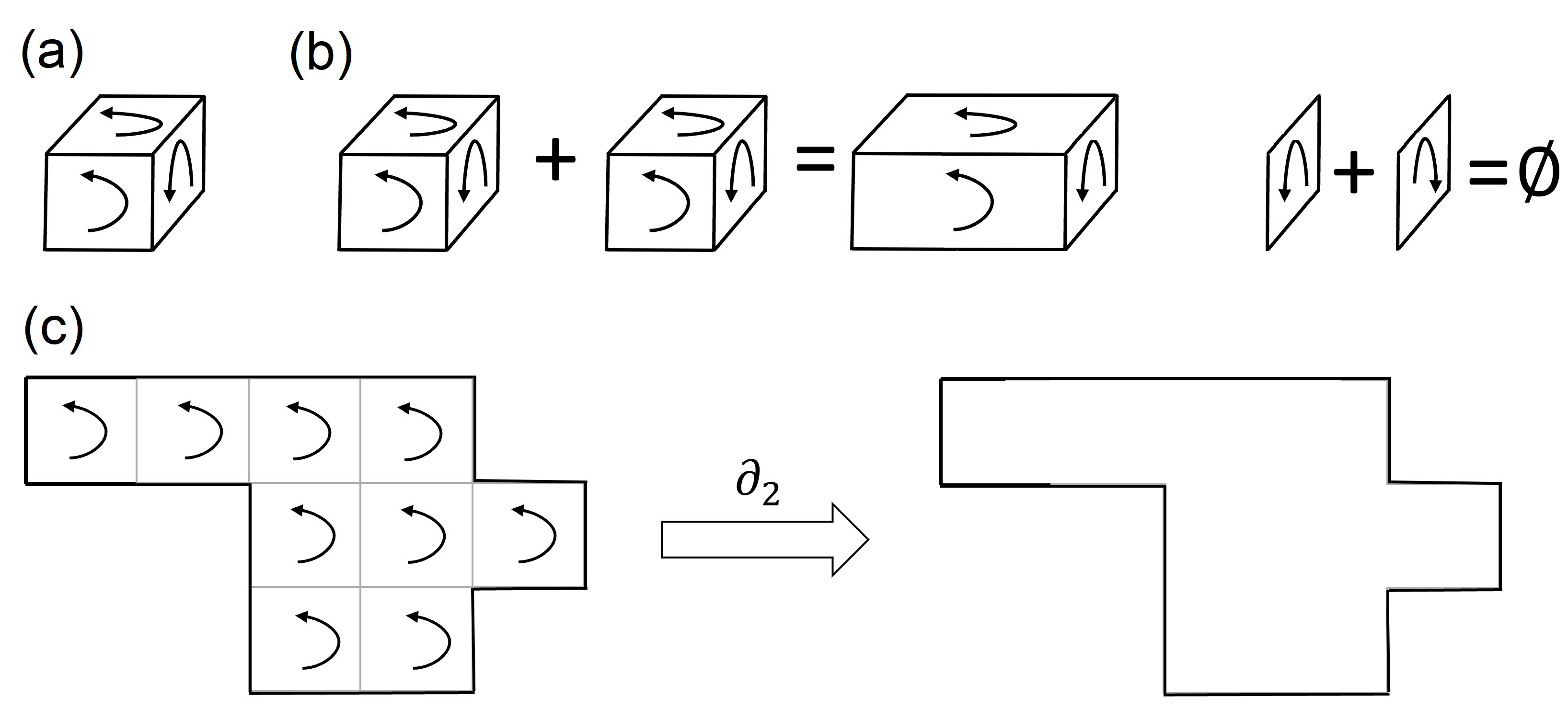}
    \caption{\sf
    An illustration of cubic complex. a). A 3-cubic cell, with the orientation of its 2-faces shown. 
    b). Two 3-cubes are summed to form a 3-chain.  The internal square is contributed twice from the two cubes.  As each surface is oriented (\textit{i.e.,} counter-clock-wise by the outward surface normal), these two squares have opposite orientations and cancel each other when summed. c). An example of a 2-chain formed by 2-cells and its boundary.
    }
    \label{fig:cubics}
\end{figure}

\paragraph{Complex and chain.}
We first discuss how to represent the $d$-dimensional configuration space $\mathbb{M}$~\cite{kaczynski2006computational}. In this study,
we use  cubic complexes~\cite{kaczynski2006computational, wagner2012efficient}.  
A $d$-dimensional cubic complex $K$ is constructed from a union of points, line segments, squares, cubes, and their $k$-dimensional counterparts glue together properly, where $k \le d$ and all have unit length (except points, which have no lengths).  We call each of these a $k$-cell or a $k$-cube (see Fig~\ref{fig:cubics}a for a 3-cell).
 While the topology of $\mathbb{M}$ is invariant whether it is represented by cubic complexes  or other complexes such as simplicial complexes, 
 the  nature of grid representation of the molecular configurations makes this choice  convenient~\cite{wagner2012efficient}.

We can build up our cubic complexe $K$ from cubes to represent the configuration spaces. 
Consider a set of $k$-cells, we can sum them up. We call the total summation of a set of $k$ cells a  \textit{$k$-chain}.
Fig~\ref{fig:cubics}b shows an example where two 3-cells are summed up to form a 3-chain.  Fig~\ref{fig:cubics}c shows how nine 2-cells are summed up to form a 2-chain. 
Here the binary operation of summing over two $k$-cells is orientation sensitive: two $k$-cells of the same underlying space but opposite orientation  cancel out each other  when summed up. 
If a set is equipped with a binary operation satisfying certain requirements, it is  called a \textit{group} mathematically.
The set of $k$-chains from the $K$ complex with our binary operation of summation therefore form a \textit{chain group} $\mathsf{C}_k(K)$.

\paragraph{Boundaries.}
We now set boundaries.
The \textit{boundary} of an individual $k$-cell  is the set of its $(k-1)$-dimensional faces, which by definition forms a $(k-1)$-chain.
The boundary of a 3-cubic cell is shown in Fig~\ref{fig:cubics}a, which is the set of the six oriented squares.
The boundary   
of a $k$-chain  is the sum of the boundaries of its element $k$-cells. Because of the nature of our sum operation, internal structures cancel out.
Consider the boundary of the two neighboring three-dimensional cubes in Fig~\ref{fig:cubics}b.  The interfacial square is contributed twice, once from each cube, but with opposite orientation as both are counter clock-wise around their outwards normals.  When these two cubes are glued together, these two boundary squares  are summed up, and they cancel each other out.  The overall outcome of this summation is indeed the outer boundary of the union of the two neighboring cubes. Fig~\ref{fig:cubics}c shows a 2-chain and its boundary.
This holds true in other dimensions as well, namely,  a $(k-1)$-dimensional face from two neighboring $k$-cells have opposite orientations and cancels each other out upon summation. 

With this summation, we  obtain the boundary of a $k$-chain  from $K$ by applying the
 \textit{boundary operator}
 $\partial_k $:
\begin{equation}
    \partial_k: 
    \mathsf{C}_k(K) \rightarrow
    \mathsf{C}_{k-1}(K).
\end{equation}

\paragraph{Cycles.}
There are certain $k$-chains 
that have no boundaries.  They are called $k$-dimensional cycles or \textit{$k$-cycles}. 
With the binary operation of summation discussed earlier, the set of $k$-cycles from $K$ form the \textit{cycle group} $\mathsf{Z}_k(K)$:
$$
\mathsf{Z}_k (K)  \equiv \{c\in \mathsf{C}_k(K)|\, \partial_k  c = \emptyset \}.
$$
As an example, we consider the three-dimensional cube again~(Fig~\ref{fig:cubics}a).  We take its six surface squares that fully enclose the solid cube.  These square form a 2-chain.  As a whole, this 2-chain itself does not have boundaries, as the six squares  are glued together along the borders and there are no openings.

\paragraph{Kernel and image.}
Analogous to the null space or kernel in linear algebra, the cycle group $\mathsf{Z}_k(K)$ is the \textit{kernel} of the operator $\partial_k $, as each of its member $k$-chain has no boundary by definition and  $\partial_k $ will send it to null:
$$
\mathsf{Z}_k (K) = \ker \mathsf{C}_k(K) \equiv \{c\in \mathsf{C}_k(K)| \partial_k c = \emptyset \}.
$$
We now move one dimension up and consider boundaries  of $(k+1)$-chains in $K$.
Boundaries are one dimension lower and therefore the boundaries of $(k+1)$-chains are $k$-chains.  They are called  the \textit{$k$-boundaries} of $K$ and form the \textit{$k$-boundary group} $\mathsf{B}_k(K)$.  As each $k$-boundary is obtained when $\partial_{k+1}$ is applied to a $(k+1)$-chain,
collectively they are the \textit{image} of
$\partial_{k+1}$:
$$
\mathsf{B}_k(K) = \text{im}\, \partial_{k+1} (\mathsf{C}_{k+1}(K))\equiv
\{c\in \mathsf{C}_k(K)| \; \partial_{k+1}(c')=c, \, c'\in \mathsf{C}_{k+1}(K) \}.
$$
It turns out that all $k$-boundaries themselves have no $(k-1)$-boundaries.
This is due to a fundamental property of the boundary homomorphisms in topology, which states that for any $k\ge 1$~\cite{munkres2018elements},
\begin{equation}
 \partial_{k-1} \circ \, \partial_{k} = \emptyset.
\label{eqn:bb=0}
\end{equation}
From our previous example in Fig~\ref{fig:cubics}a,  the 2-chain of the six squares that  enclose the solid cube is the boundary of a 3-chain (a lone 3-cell in this case).  They themselves do not have any opening, and hence the boundary of this 2-chain is $\emptyset$.
A consequence of this general property is that we have 
$\mathsf{B}_k(K) \subseteq \mathsf{Z}_k(K) \subseteq \mathsf{C}_k(K)$.

\begin{figure}[h!]
    \centering
    \includegraphics[scale=0.7]{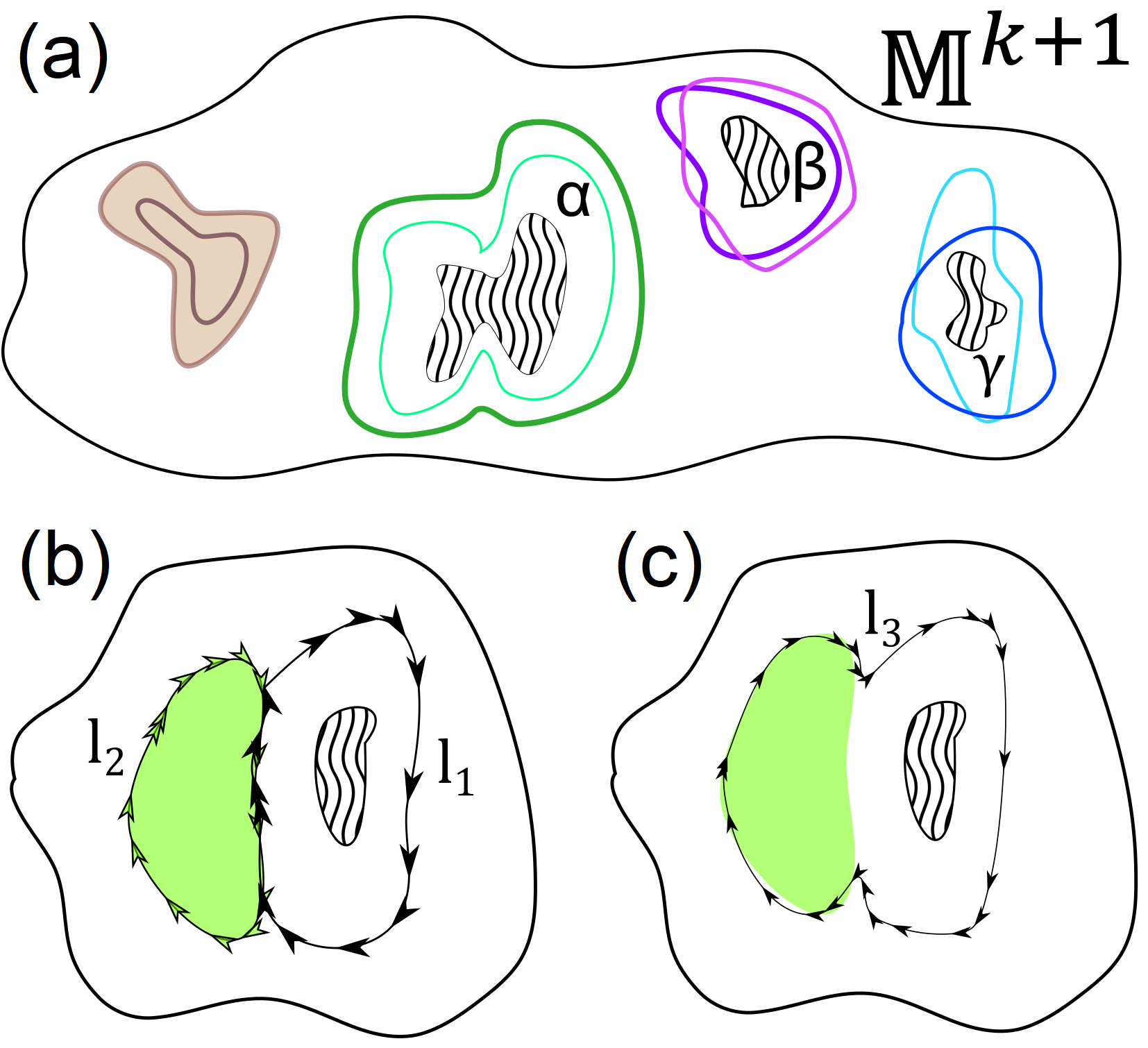}
    \caption{\sf
    An illustration of homology classes of $k$-cycles in a $(k+1)$-manifold $\mathbb{M}_{k+1}$. 
    a). The two $k$-cycles in brown and dark brown enclose a $(k+1)$-body ($k$-chains  in light brown). They contain no holes and are boundary cycles. They belong to the same equivalence class of $[\emptyset]$.
    The $k$-cycles in green/light green, purple/light purple, and blue/light blue each contain  $k$-holes $h_\alpha,\, h_\beta$ and $h_\gamma$, respectively, and are part of equivalence classes of $[h_\alpha],\,[h_\beta]$ and $[h_\gamma]$, respectively.
    b).
    The $k$-cycle $l_1$ contains a hole. The boundary cycle $l_2$ contains no hole but a $(k+1)$ body. 
    Note that $l_1$ and $l_2$  share a common piece of boundary but in opposite orientations.
    c). When  $l_1$ and $l_2$ are summed up, we obtain the $k$-cycle $l_3$, which contains the same hole as $l_1$.  Both $l_1$ in b) and $l_3$ in c) belong to the same equivalence class of $k$-cycles containing this hole. 
    }
    \label{fig:loops}
\end{figure}
\paragraph{Homology group and Betti number.}
There are two types of $k$-cycles:  Those enclose $(k+1)$-bodies and those enclose $(k+1)$-holes~(Fig~\ref{fig:loops}a).  The former are boundaries of the enclosed bodies of $(k+1)$-chains and can be collapsed into a point (Fig~\ref{fig:loops}a, $k$-cycles in dark brown/brown enclosing $(k+1)$-chains shaded in light brown ).  The latter are not boundaries of  $(k+1)$ bodies and cannot be collapsed into a point (Fig~\ref{fig:loops}a, other cycles) .  We will first distinguish these two types of cycles. 
Furthermore, among cycles that do not enclose a body, they may be so due to different reasons, as the holes they contain may be different (Fig~\ref{fig:loops}a).  We will distinguish these different situations as well. 

We consider all cycles containing the same hole essentially the same and group them into one \textit{equivalence class} of cycles.  As an illustration, green/light green $k$-cycles in Fig~\ref{fig:loops}a form an equivalence class $[h_\alpha]$ as they all contain hole $h_\alpha$.  
So do the purple/light purple $k$-cycles (class $[h_\beta]$ containing hole $h_\beta$), and the blue/light blue $k$-cycles (class $[h_\gamma]$ containing hole $h_\gamma$).
A special equivalence class are cycles containing the $\emptyset$ hole or no hole (Fig~\ref{fig:loops}a, brown/dark brown cycles, class $[\emptyset]$).  
We call  cycles  in each equivalence class \textit{homologous} to each other. If they encircle different holes, they belong to different equivalence classes.

We elaborate on this. Among all elements of $\mathsf{Z}_k(K)$, which are $k$-cycles, we identify  all $k$-boundaries, which contain  $(k+1)$ bodies and are elements of $\mathsf{B}_k(K)$. Because of Eqn~(\ref{eqn:bb=0}), they have no $(k+1)$-holes. We put them into a class denoted as $[\emptyset]$ as they contain no holes (or $\emptyset$-hole) (Fig~\ref{fig:loops}a, brown/dark brown cycles $\in [\emptyset]$). For the remaining $k$-cycles of $\mathsf{Z}_k(K)$, they are not in $[\emptyset]$ but  may contain different holes. We identify those contain hole $h_a$, and put them into the equivalence class denoted as $[h_a]$ (Fig~\ref{fig:loops}a, green/light green cycles $\in [h_a]$).  Remaining $k$-cycles that contain hole $h_b$ are put into the class $[h_b]$, and so on (Fig~\ref{fig:loops}a, purple/light purple cycles $\in [h_b]$, and blue/light blue cycles $\in [h_c]$).  Each element of the set $\{[\emptyset],\,[h_a],\, [h_b],\,\cdots\}$ is an \textit{equivalence class}.

As these equivalence classes themselves form a set, and the outcome of the binary operation of summation on elements of $\mathsf{Z}_k$ is preserved, this set form a new group. This new group is called a \textit{quotient group}, as it is obtained from $\mathsf{Z}_k(K)$ after factoring out the boundaries $\mathsf{B}_k(K)$.
The \textit{ $k$-th homology group} $\mathsf{H}_k(K)$ is this quotient group:
$$
\mathsf{H}_k(K) \equiv \mathsf{Z}_k(K)/\mathsf{B}_k(K).
$$
The elements of $\mathsf{H}_k(K)$ are equivalence classes of homologous cycles representing the holes (or lack of) they encloses.

Two $k$-cycles are homologous to each other if they contain the same hole, or equivalently, if one can be obtained from another by adding  a $k$-boundary~(Fig~\ref{fig:loops}b and Fig~\ref{fig:loops}c). To illustrate this, note that the cycle labeled $l_3$ in Fig~\ref{fig:loops}c can be obtained by adding the $k$-boundary  of a $(k+1)$-body  labeled $l_2$ to the  $k$-cycle labeled $l_1$ enclosing a hole (Fig\ref{fig:loops}b). Due the cancellation nature of our summation, the commonly shared piece of the boundaries is cancelled out and we have the larger $k$-cycle $l_3$ in  Fig~\ref{fig:loops}c enclosong the same hole.  Upto the difference of a boundary of a solid $(k+1)$-body, these two $k$-cycles are the same and belong to the same equivalence class.  It is not difficult to see that we can repeat this operations of adding certain $k$-boundaries and convert any homologous $k$-cyles between one another.

The number of the equivalent classes, or the number of independent $k$-dimensional holes, is counted by the dimension of the homology group.  It is called the \textit{$k$-th Betti number} $\beta_k(K)$:
$$
\beta_k(K) = \dim (\mathsf{H}_k(K) )
$$

\begin{figure}[h!]
    \centering
    \includegraphics[width=0.7\textwidth]{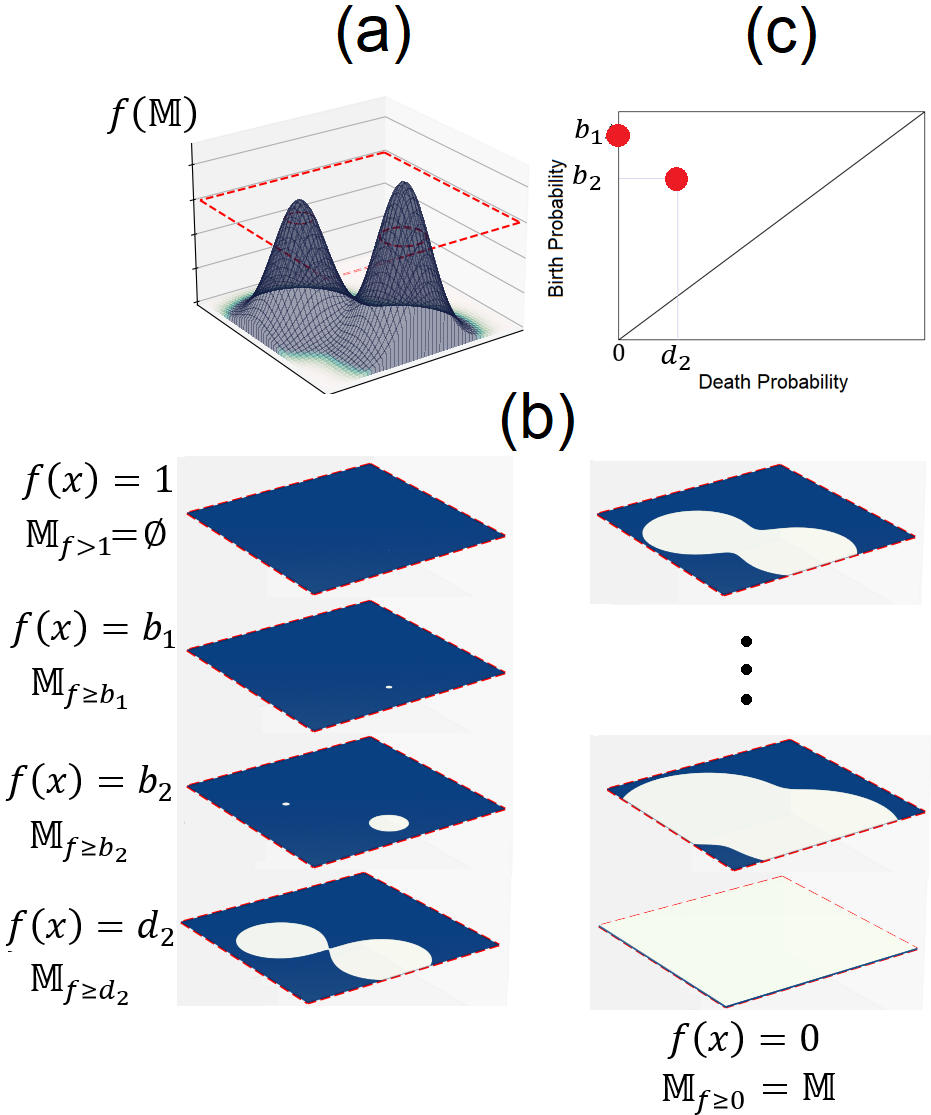}
    \caption{\sf
  The probability landscape $f(\mathbb{M})$ and the topology of its superlevel set $\mathbb{M}_{f\ge }$. a) The landscape and a sea level. The superlevel sets $\mathbb{M}_{f\ge}$ are the regions of the domain $\mathbb{M}\subset \mathbb{\real^d}$ (shown as a plane) whose landscape value is above the sea. b) At $f(\bx)=1$, all is below the sea level and $\mathbb{M}_{f\ge1}=\emptyset$.
  At $f(\bx)=b_1,\, b_2$ and $d_2$, the topology of $\mathbb{M}_{f\ge }$ (shown in white) changes. 
  At $f(\bx)=0$, all is above sea level and we have $\mathbb{M}_{f\ge } = \mathbb{M}$.
  c) The persistent diagram of the birth and death value of $f(\bx)$  for the $0$-th homology group representing the two peaks. 
  The sublevel sets  below the sea $\mathbb{M}_{f<}$ are shown in blue.
   }
    \label{fig:topoLevels}
\end{figure}

\paragraph{Filtration.}
We now examine the topological structures of holes in the probability landscape on the configuration space,  when we restrict  to configurations all with probabilities above certain value.  By gradually adjusting this value, we will be  able to trace out the details of topological changes.
For an illustration, envision a sea on top of the probability landscape~(Fig~\ref{fig:topoLevels}). At the level of $f(\bx) = 1$, it covers the whole landscape.  The domain of the part of the landscape  above the sea level is $\emptyset$.
We gradually lower the sea level to value $b_1$, when the first peak emerges from the sea (birth of the first peak). At this time, we have the superlevel set $\mathbb{M}_{f\ge b_1}$, which are the set of points $\{\bx \in \mathbb{M}| f(\bx) \ge b_1 \}$. They form the white region(s) in Fig~\ref{fig:topoLevels}.  We further lower the sea level to $b_2$  when another peak emerges (birth of the second peak), at which time we have the superlevel set $\mathbb{M}_{f\ge b_2}$.
Suppose we continue this process until sea level reaches $d_2$  where the two peaks are merged together (death place of the second peak) by a land ridge  that has just emerged above the sea level.  At this sea level, we have $\mathbb{M}_{f\ge d_2}$. At each of these levels, the topology of the superlevel set changes, namely, one component, two components, and then one component again.  These changes are captured by the changing homology groups and the Betti numbers.

We now generalize.  We have a descending sequence of probability values corresponding to the lowering sea level:
$$
1=a_0>a_1 > a_2 > \cdots > a_n = 0,
$$
and the corresponding superlevel sets, or the domains of the part of the landscape above the sea level, which are subspaces of $\mathbb{M}$:
$$
\emptyset =\mathbb{M}_0 
\subset \mathbb{M}_1 
\subset \mathbb{M}_2 
\cdots
\subset \mathbb{M}_n = \mathbb{M}. 
$$
Recall we have the full configuration space $\mathbb{M}$ represented by a cubic complex $K$. Each superlevel set $\mathbb{M}_i$ is represented by a subcomplex $K_i \subset K$, which can be derived from the original full complex $K$. We then have the corresponding sequence of subcomplexes:
$$
\emptyset = K_0
\subset K_1 \subset K_2 \cdots \subset K_n = K.
$$
This sequence of subcomlexes is called a \textit{filtration}.  

We are interested in how the topology of $\mathbb{M}_{f\ge a_i}$ evolves at different $a_i, i = 0, \cdots, n$. This is represented by the corresponding sequence of homology groups connected by linear maps:
$$
0= \mathsf{H}_k(K_0)\rightarrow
\mathsf{H}_k(K_1)\rightarrow \cdots
\rightarrow \mathsf{H}_k(K_n)= \mathsf{H}_k(K).
$$

\paragraph{Persistence and Persistent diagram.}
As we move from $K_{i-1}$ to $K_i$, we may gain a new equivalence class,
(\textit{e.g.}, a new peak for 
$0$-th homology as in our example), 
or we may lose one 
(\textit{e.g.} when a peak is merged with another one). We say that an equivalence class of a $k$-cycle $[\alpha_i]$ is \textit{born} at $a_i$ if its equivalence class is present in  $K_i$ but absent in $K_{i-1}$ for any value of $a_{i-1} < a_i$.  The class 
\textit{dies} at $a_i$  if it 
is present in $K_{i-1}$ for any value of $a_{i-1} < a_i$ but not at $a_i$.  
We record the location and the value of $a_i$,  namely, the corresponding $k$-cube and its probability value whose inclusion lead to the birth 
and death 
events. 
 
The prominence of the topological feature of a $k$-cycle is encoded in its life-time or \textit{persistence}. Denote the birth value  and the death value of class $[\alpha_i]$ as $b_i$ and $d_i$, respectively.
The \textit{persistence} of class $[\alpha_i]$ is then $b_i-d_i$.

In the example shown in Fig~\ref{fig:topoLevels}, the equivalence class of $0$-cycles (components) associated with the first peak is born at $f(\bx)=b_1$. The equivalent class associated with the second peak is born at $f(\bx)=b_2$.
At $f(\bx)=d_2$, these two components merge together. We say that the second peak  dies at $d_2$, and its persistence is $b_2-d_2$.  The first peak dies at $f(\bx)=0$, and its persistence is $b_1-0=b_1$.

We record the birth and death events of  homology classes in a two-dimensional plot, which is called the \textit{persistent diagram}. 
Each homology class is represented by a point in this diagram, where the birth value $b_i$ and the death value $d_i$  are taken as its coordinates ($b_i,\, d_i$). Fig~\ref{fig:topoLevels}c shows the persistent diagram of our illustrative example.

In general, we have the $k$-th persistent diagram $\mathcal{P}_k(f)$ of $k$-cycles  for  our probability function $f: \mathbb{M} \rightarrow \real_{[0,1]}$. It is the set of points such that each point $(x_1, x_2)$ represents a distinct topological feature of $k$-cycle, which is present in 
$ \mathsf{H}_k(\mathbb{M}_{f\ge a})=
\mathsf{H}_k(f^{-1}([a, 1))$
for $a\in [x_1, x_2)$.

\paragraph{Computation.}
The key to study homology groups in high-dimensional space is the construction of the $K$-complex to represent the configuration space $\mathbb{M}$. 
In this study, we use the cubic algorithm of~\cite{wagner2012efficient} with modifications~\cite{Tian-topo-preprint}, so it can be applied to higher dimensions.  We consider  only the $0$-th persistent homology groups, which records the birth and death of probability peaks. The locations $\bx$ where birth and death events occur, namely, the corresponding $k$-cubes are also computed.

\section {Results}

\subsection{Model system and computation}
\paragraph{Ensemble of reactive trajectories and conformations.} 
The isomerization of alanine dipeptide in vacuum provides a  tractable system for understanding the process of activation in details, and has been well studied as a model for  understanding protein conformational changes~\cite{Ma2005Automatic,Li_Ma2016TPS,2000Bolhuis1}.

\begin{figure}[h!]
    \centering
    \includegraphics[width=0.99\textwidth]{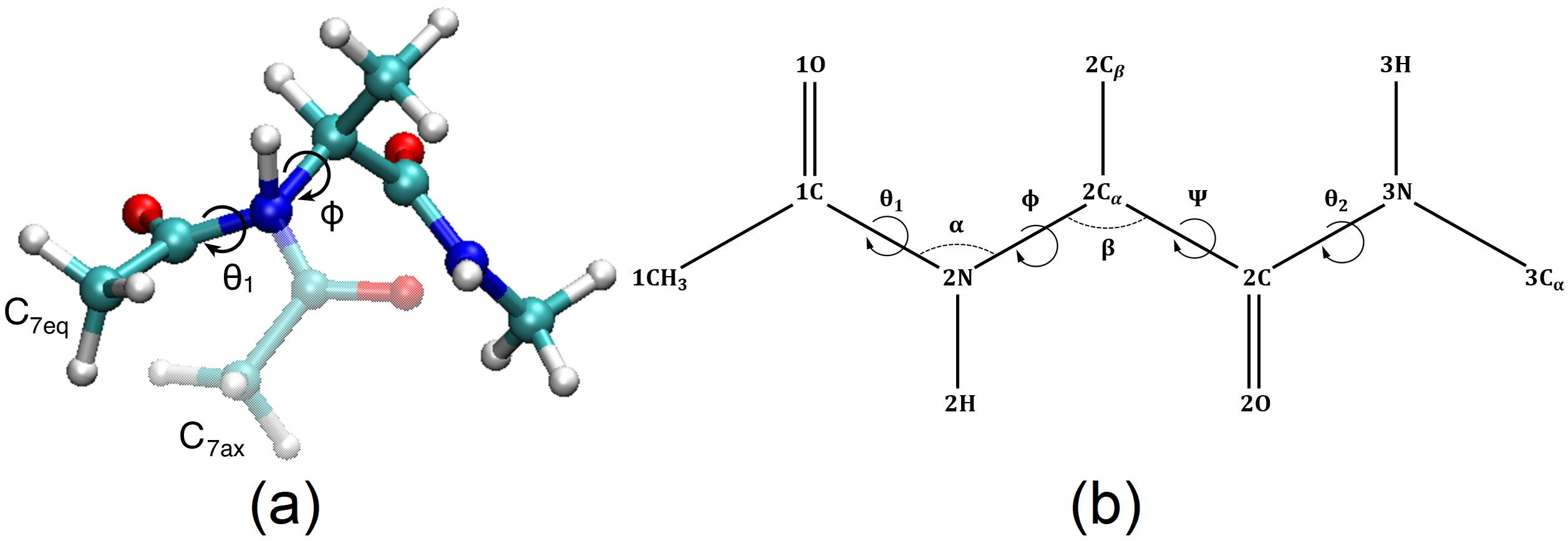}
    \caption{\sf The isomerization reaction of alanine dipeptide. (a) Conformations from the reactant and product basins before and after the isomerization.
    (b) The six reaction coordinates of the isomerization process of alanine peptide examined in this study.}
    \label{fig:alanine}
\end{figure}

Using transition path sampling~\cite{Bolhuis2002}, we harvested 6 million reactive trajectories. 
Each trajectory is of 2.5 ps duration.  We further collect conformations every 50 steps at 1 fs/step along each trajectory. 
Altogether, we have a total of 
$1.5 \times 10^{11}$ conformations. 
All simulations are conducted using the molecular dynamics software suite GROMACS4.5.4~\cite{Berk2008}. Amber$94$ force field was used to facilitate the comparison with previous results. The simulation was performed with constant total energy $36$ KJ/mol, such that the averaged temperature is $300$K for the transition path ensemble. 
Note that the transition portion of each reactive trajectory is around $0.2$ ps, thus the majority  of our $2.5$ ps trajectories are within the two stable basins.
Here the reactant basin  is defined in radian as
$ (\phi,\, \psi) \in 
[(-3.49,\,-0.96)\times (-1.57,\,3.32)]$ 
and the product basin  is defined as
$ (\phi,\, \psi) \in 
[(0.87,\,1.74)\times (-1.39,\,0)]$. 

\paragraph{Constructing dynamic probability surface of transition state.}
We then construct the dynamic probability surface of the isomeriztion reaction from the sampled $1.5\times 10^{11}$ conformations.  
With a balanced consideration of the available MD simulation trajectories and the dimensionality, we construct a 5-dimensional  configuration space for this study.
Based on previous analysis using the energy flow theory~\cite{Li_Ma2016_Reaction_Mechanism}, we  selected the top 5 coordinates ($\phi, \theta_1, \psi, \alpha, \beta$) that contribute most to the activation dynamics.
The original 60 dimensional  space is then projected onto this 5 dimensional space, where each dimension is divided into 15 bins.
This leads to $15^5=759,375$ 5-dimensional hypercubes.

\paragraph{Computing topological structures of the dynamic probability landscape.}
We then carry out  persistent homology analysis. Computations are conducted on a machine with a 20-core Xeon E5-2670CPU of 2.5 GHz, with a cache size of 20 MB and memory of 128 GB Ram. The computing time for finding the significant peaks and  ridges connecting them is $\approx$ 30 seconds.

\paragraph{Committor test for conformations at selected locations of the configuration space.} 
We carry out committor test for configurations of the dipeptide identified by persistent homology.
The committor value of a configuration is defined as the probability that a dynamic trajectory initialed from this configuration, with initial momenta drawn from the Boltzmann distribution, reaches the product basin before the reactant basin. A configuration with committor value $p_B=0.5$ is regarded as a member of the transition state ensemble.

In a committor test~\cite{Bolhuis2002,Du1998}, we need to generate an ensemble of tentative transition state conformations from locations in configurtaion space where probability peaks and ridges identified by persistent homology  are located.
 These conformations all share the same target values for the selected coordinates (\textit{e.g.}, $\phi$, \,$\theta_1$) that correspond to the location of the selected peak, with the other coordinates sampling the equilibrium distribution.  
For this, we add harmonic restraint potentials on the selected coordinates to the system potential energy function.  The minima of the harmonic restraints are at the target values.  Equilibrium MD simulations are then carried out.  Conformations harvested from such simulations are filtered to generate an ensemble of conformations that all share the same target values for the selected coordinates.  The restraint potential is used to enrich conformations that satisfy this criterion.

\subsection{Topology and dynamic properties of 5-d dynamic probability surfaces}

\begin{figure}[h!]
\centering
\includegraphics[width=0.99\textwidth]{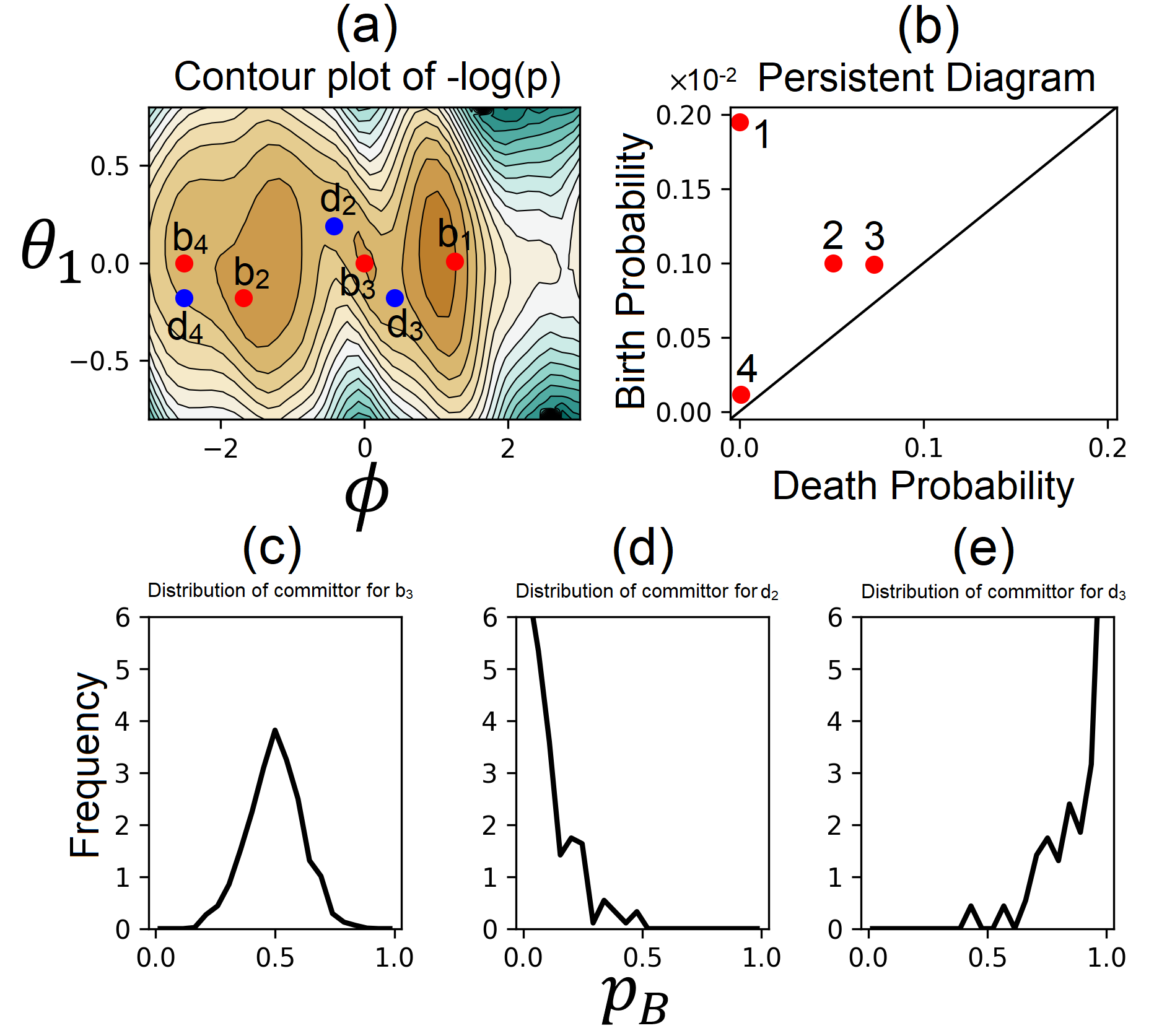}
\label{fig:sub1}
\caption{\sf The 5-d dynamic probability surface on the ($\phi-\theta_1$) plane, its topological structures, and the committor values.
(a) The 5-d dynamic probability surface $p(\phi, \psi, \theta_1, \alpha, \beta)$ shown on the ($\phi-\theta_1$) plane. 
Red and blue dots are locations of probability peaks and ridges (see also~SI Table~S1).
(b) The persistent diagram 
recording the birth and death probabilities $p(b_i)$  and $p(d_i)$  of the peaks in $y$ and $x$, respectively.
(c-e)  Distributions of committor values $p_B$  for trajectories from locations of $b_3$, $d_2$, and $d_3$, respectively. 
(c) The transition state ensemble is located at $b_3$.
}
\label{fig:fivetrue}
\end{figure}

\paragraph{Dynamic probability surface on configuration space of ($\phi,\, \theta_1,\, \psi, \alpha, \beta$).}
We examine the topological structures of this 5-d dynamic probability surface.
There are four peaks, each located in a 5-d cube (Fig.~\ref{fig:fivetrue}a, red dots for birth locations of peaks and blue dots for their ridges or death locations, and values are listed in SI Table~S1).
The most prominent peak with the largest persistence is peak $b_1$ (see persistent diagram  in Fig.~\ref{fig:fivetrue}b), which corresponds to the product basin.
The second most prominent $b_2$ is the reactant basin.
The third prominent peak at $b_3$ (Fig~\ref{fig:fivetrue-3D}a)
 has roughly the same  probability as the reactant basin  $b_2$, but a shorter persistence, namely, it does not stand out from the surrounding landscape as much. 
It subsequently merges with the peak at the product basin $b_1$.
As expected, peaks have all negative eigenvalues for their Hessian matrices, and ridges have one positive and four negative eigenvalues (see Supplementary Info).


We then  take conformations from  the four  peaks 
and the three ridges~(SI Table~S1),
and carry out committor tests.
We find that the most prominent peak $b_3$ beyond the reactant and product basins fully captures the transition state ensemble  ($p_B$ centered at 0.5, Fig~\ref{fig:fivetrue}c):  Trajectories initiated from configurations at this location have equal probability towards the reactant or the product basin.

In contrast, all committor values $p_B$ for locations of $b_2$, $d_2$, and $d_4$ are found to be 0.  Reaction trajectories starting from conformations at these locations fall back to
the reactant basin. 
The committor values for  peak $b_1$ are all 1.0: Trajectories from this location all go to the product basin. 
The committor values for conformations at the ridges $d_2$ and $d_3$ follow one-sided distributions (Fig~\ref{fig:fivetrue}d-\ref{fig:fivetrue}e, respectively). 
Only a negligible amount of conformations have $p_B = 0.5$. 

These results demonstrate that the dynamic properties of the transition state ensemble is capture by the topological features of this dynamic probability surface.  Furthermore,   the transition state ensemble is located at the most prominent peak outside of the reactant and product basins, instead of a saddle point.

\begin{figure}[h!]
\centering
\includegraphics[width=0.99\textwidth]{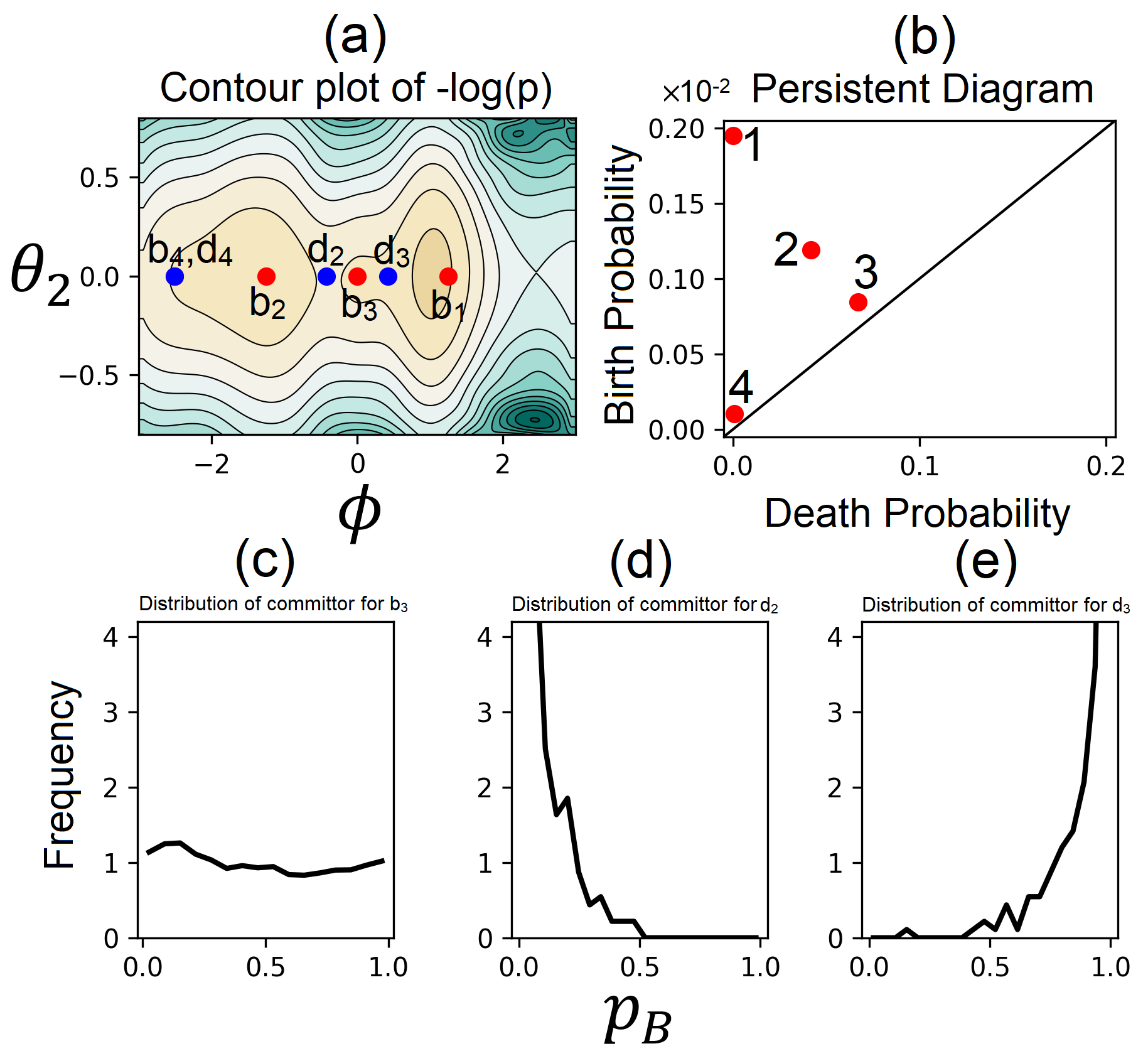}
\caption{\sf A different 5-d dynamic probability surface with $\theta_2$ replacing $\theta_1$ on the ($\phi-\theta_2$) plane, its topological structure represented in persistent diagram, and distributions of committor values. (a) The 5-d dynamic probability surface $p(\phi, \psi, \alpha, \beta, \theta_2)$ projected onto the ($\phi-\theta_2$) plane.
Red and blue dots are locations of probability peaks and ridges. 
(b) The persistent diagram 
recording the birth and death probabilities $p(b_i)$  and $p(d_i)$  of the peaks in $y$ and $x$, respectively.
 (c-e) Distribution of the committor values $p_B$
 for trajectories from locations of $b_3$, $d_2$, and $d_3$.
}
\label{fig:fivenew}
\end{figure}

\paragraph{Dynamic probability surface on configuration space ($\phi,\,  \psi, \alpha, \beta,\, \theta_2$).}
To examine the importance of proper choice of the coordinates, we construct another 5-d probability surface by omitting the coordinate $\theta_1$ and replacing it with $\theta_2$, which is on the other end of the molecule in a position symmetric to $\theta_1$.
Fig.~\ref{fig:fivenew}a shows the projection of the 5-d surface in $-\ln p(\bx)$ on the ($\phi-\theta_2$) plane.

There are four significant  peaks (Fig.~\ref{fig:fivenew}a, red and blue dots), which are also shown in the persistent diagram (Figure~\ref{fig:fivenew}b),  each located in a 5-d cube (see SI Table~S1). 
Peaks $b_1$ and $b_2$ are the most and second most prominent peaks,  
 corresponding to the product and the reactant basins, respectively. 
In this projection, peaks are separated only in $\phi$ and they have almost identical $\theta_2$ values. This differs from the 5-d surface containing $\theta_1$  (Fig~\ref{fig:fivetrue}). 

We then sample conformations from the location of each of the  
four peaks and the three ridges~(SI Table~S1) and perform committor tests. 
The committor values for  peak $b_1$ are all  1.0, where trajectories starting here all go to the product basin. 
All committor values for conformations at  $d_2$, $b_4$, and $d_4$ are 0.0. Reaction trajectories  from  these locations all go to the reactant basin. 
The committor values for $d_2$ and $d_3$  follow one-sided distributions (Fig.~\ref{fig:fivenew}d-\ref{fig:fivenew}e). Only a tiny amount of the conformations have $p_B=0.5$, indicating that the  transition state ensemble are  located elsewhere. 

The committor values for $b_3$ has a flat distribution. While there are conformations with $p_B$ values close to 0.5,  their frequency is similar to any other $p_B$ values.  This indicates that the 5-d cube where $b_3$ is located contains some transition state conformations as its $\phi$ value is correct, but also a mixture of other conformations with diverse dynamic properties.
Overall,  without  the reaction coordinate $\theta_1$, this 5-d dynamic probability surface does not
describe the activated process adequately,  and cannot be used to identify the transition state ensemble.

\subsection{Topology and dynamic properties of projected 2-d dynamic probability surfaces.}  
\begin{figure}[h!]
    \centering
    \includegraphics[width=0.99\textwidth]{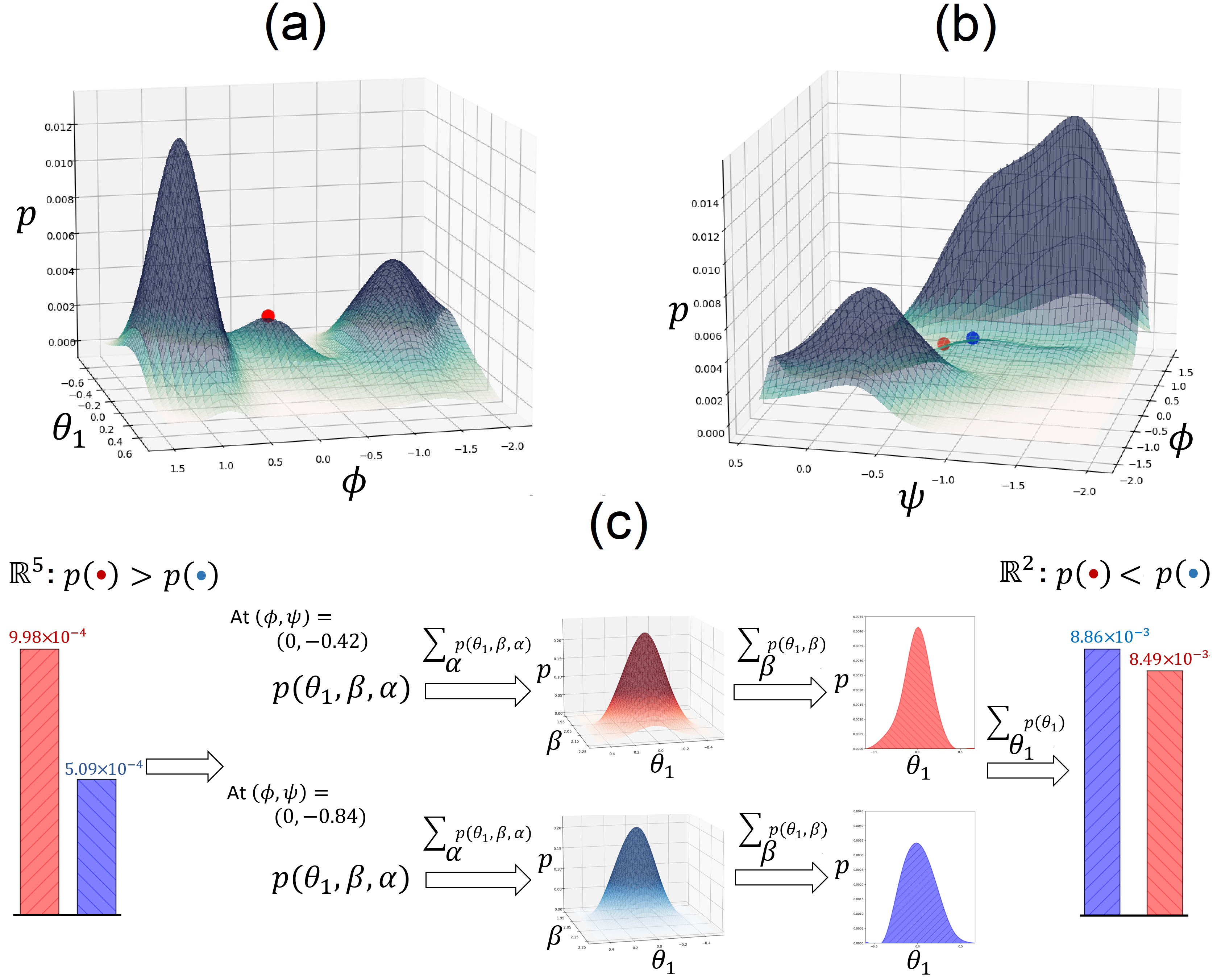}
    \caption{\sf  The 5-d probability surface projected onto two different 2-d planes. (a) The probability surface projected onto the ($\phi-\theta_1$) plane. The red dot is the location of the third probability peak after reactant and product peaks, which is  where the transition state conformations are located.
    (b) The probability surface projected onto the ($\phi-\psi$) plane. 
    Conformations with the correct transition state value of $\phi$ are at a non-peak location, which is on a slope  below the new peak shown in blue.
    (c) The process of projecting the 5-d probability surface onto the 2-d ($\phi-\psi$) plane for the blue and red dots shown in (b). While the probability at the location of the red dot is larger on the 5-d probability surface (left bars), after projection onto the  ($\phi-\psi$) plane, the probability at the red dot is smaller than that at the blue dot (right bar plots). As a consequence, the projected probability surface on the ($\phi-\psi$) plane does not capture the actual location of the peak in the 5-d surface.}
    \label{fig:fivetrue-3D}
\end{figure}

As it is difficult to directly study  the topology of a high-dimensional surface, a common practice is to project the surface to a lower dimensional subspace and  analyze the  topology of the projected surface instead.  The caveat of this practice is that the  original topological features may be lost,  new features that are artifacts may arise due to the marginalization of the probability distributions (See Fig.~\ref{fig:fivetrue-3D} for details).

To assess how well the dynamic properties is retained in a subspace, 
we project the 5-d  surface on ($\phi$, $\theta_1$, $\psi$, $\alpha$, $\beta$)
to 2-d planes. We then analyze the topological structures of the projected surfaces, and assess the dynamic behavior of the identified topological features.
We carried out this analysis using the 2-d planes of ($\phi-\theta_1$) and
($\phi-\psi$).
These choices are based on prior knowledge of the role of $\phi$ and $\theta_1$ in alanine-dipeptide, and the widely practiced ($\phi-\psi$) based Ramachandran plot.

\begin{figure}[h!]
\centering
\includegraphics[width=0.99\textwidth]{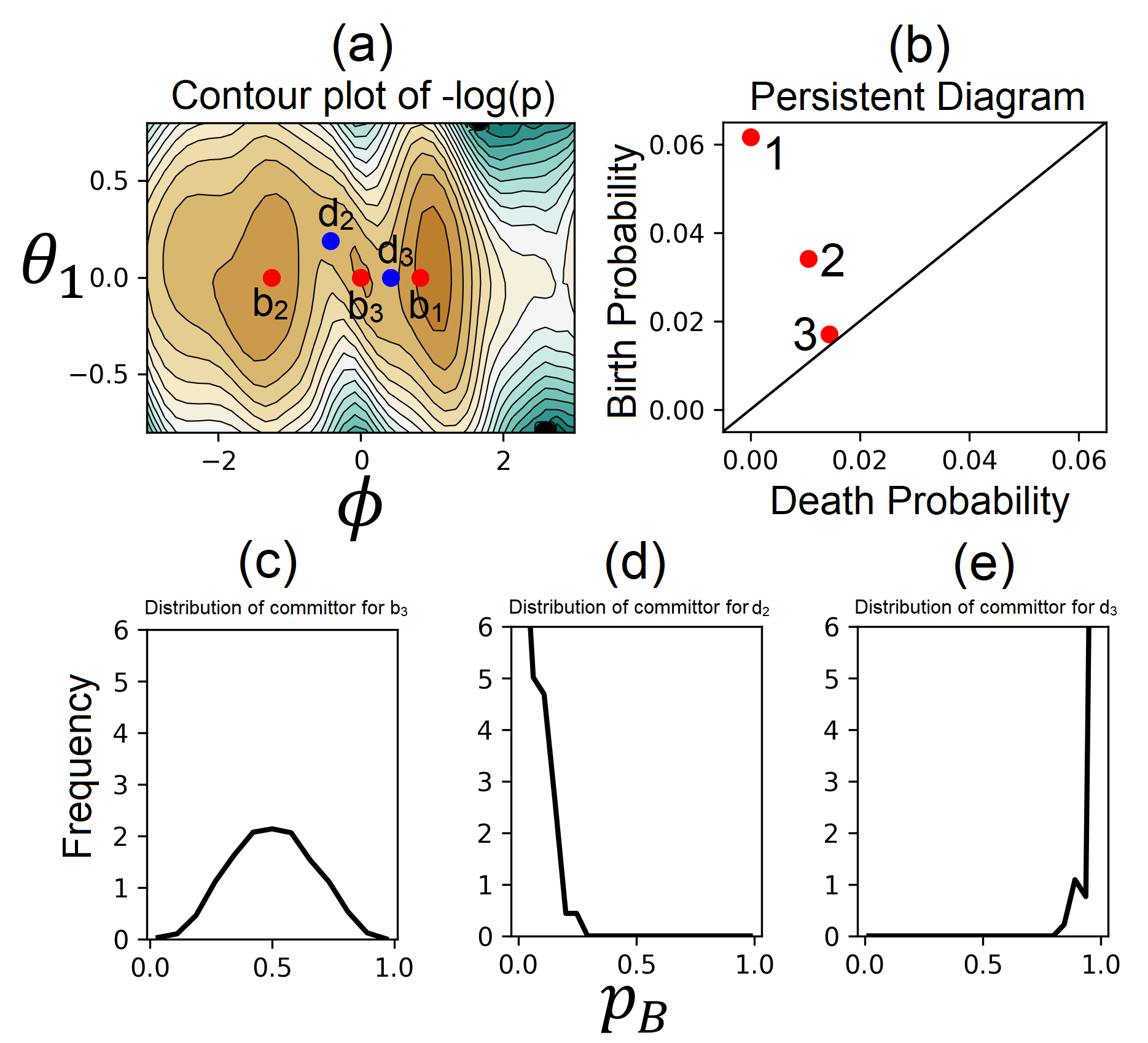}
\caption{\sf The ($\phi$-$\theta_1$)-projection of the 5-d dynamic probability surface, its topological structure, and distributions of committor values. (a) The 5-d dynamic probability surface projected onto the ($\phi-\theta_1$) plane. Red and blue dots are locations of probability  peaks  and  ridges, respectively. (b) The  persistent  diagram   recording the  birth  and  death  probabilities $p(b_i)$ and $p(d_i)$ of  the  peaks  in  $y$ and$x$,  respectively.
 (c-e) Distributions of committor values $p_B$ 
 for trajectories from peaks and ridges $b_3$, $d_2$, and $d_3$, respectively. (c)  Transition state conformations are at $b_3$.
}
\label{fig:marginalized}
\end{figure}

\paragraph{Projecting 5-d dynamic probability surface  to the ($\phi-\theta_1$) plane.}
After projection,  
there are only three significant probability peaks 
(Fig.~\ref{fig:marginalized}a, Fig~\ref{fig:fivetrue-3D}a,  red and blue dots, and SI Table~2),  
instead of four peaks found in the full $5$-d probability surface of Fig.~\ref{fig:fivenew}.
The most  and the next prominent peaks $b_1$ and $b_2$  shown in the persistent diagram of Fig.~\ref{fig:marginalized}b
are the product and reactant basins, respectively.

It is informative to 
compare peak locations on the 5-d  surface  
to that on the projected  surface  (SI Table~S1 and SI Table~S2).
The $\phi$ coordinate 
for the product basin $b_1$ is altered from 1.25 to 0.84 after projection, while the $\theta_1$ coordinate is unchanged. 
The $(\phi,\,\theta_1)$ coordinates of  the reactant basin $b_2$ are also changed from $(-1.68, \, -0.18)$  to $(-1.25,\, 0.0)$.  
$\theta_1$  of the ridge $d_3$ is changed from $-0.18$ to $0.0$,
while $\phi$ is unchanged.
The fourth prominent peak becomes undetectable after projection. 
The  persistence diagram (Fig.~\ref{fig:marginalized}b) is also significantly different: The third peak is much less prominent with reduced  persistence compared to that in Fig.~\ref{fig:fivetrue}b  (see also SI Table~S2).

We then carry out committor test on conformations from locations of these topological features.
The distribution of $p_B$ sampled from trajectories from $b_3$ is centered around 0.5 (Fig~\ref{fig:marginalized}c) and exhibits significant enrichment of
 transition state conformations.
This is similar to peak $b_3$ on the original surface over the 5-d configuration space~(Fig~\ref{fig:fivetrue}c),   
 although the distribution has a broader width.  
 
The committor values of trajectories  from the product basin $b_1$ are  all 1.0, as they all fall back to the product basin. Similarly, trajectories from  $b_2$ all fall back to the reactant basin with a $p_B$ value of 0.0.
The committor values for the bridges at $d_2$ and $d_3$  follow one sided distributions, but only a small amount of conformations have $p_B = 0.5$ (Fig.~\ref{fig:marginalized})d and e.

Overall, these results demonstrate that when projected to the  2d-plane of ($\phi$-$\theta_1$), which is formed by the two dominant reaction coordinates, the dynamic probability surface retain essential dynamic properties of the transition state surface, and contain rich information such that the transition state conformations can be recovered.

\begin{figure}[h!]

\includegraphics[width=0.99\textwidth]{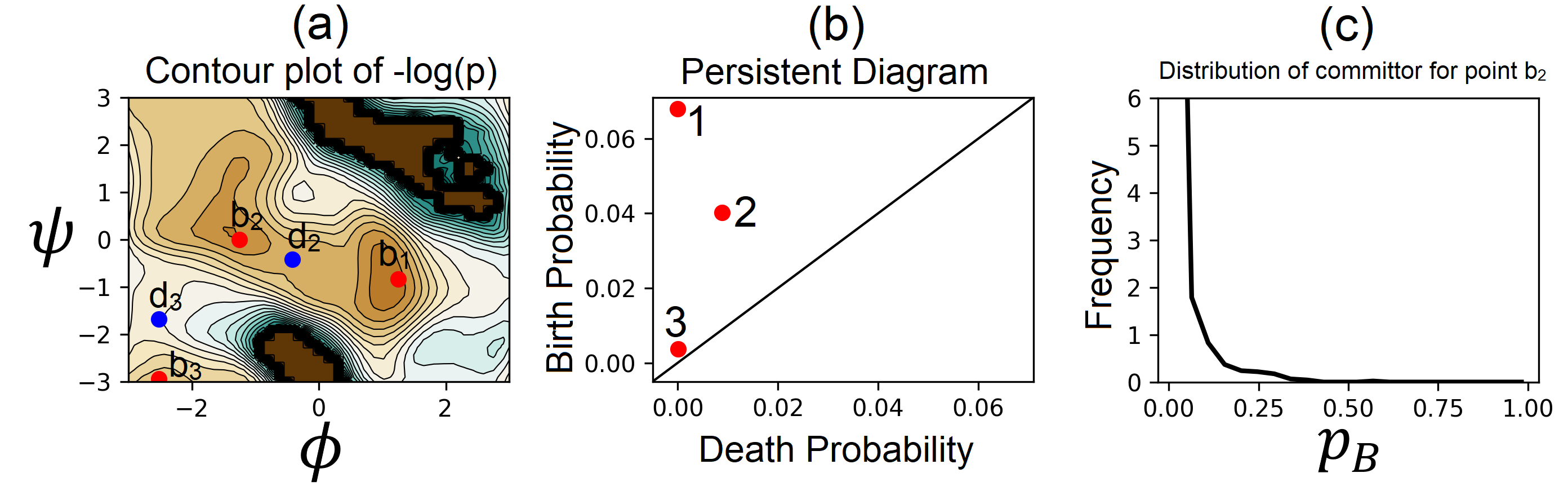}
\caption{\sf The ($\phi$-$\psi$)-projection of the 5-d dynamic probability surface, its topological structure, and distribution of committor values. 
(a) The 5-d dynamic probability surface projected onto the ($\phi$-$\psi$) plane. Red and blue dots are  locations of probability peaks and ridges, respectively.
(b) The persistent diagram recording the birth and death probabilities $p(b_i)$  and $p(d_i)$  of the peaks in $y$ and $x$, respectively.
(c) Distribution of committor values $p_B$ for peak $b_2$. }
\label{fig:margialized_phi-psi}
\end{figure}

\paragraph{Projecting 5-d dynamic probability surface to the ($\phi-\psi$) plane.}
$\phi$ and $\psi$ angles are the standard parameters to describe protein secondary structures. 
After  projection to the ($\phi-\psi$) plane,   
there are three significant probability peaks (Fig.~\ref{fig:margialized_phi-psi}a, red and blue dots, and  SI Table~S2).
The most  and the next most prominent peaks $b_1$ and $b_2$  as shown in the persistent diagram  (Fig.~\ref{fig:margialized_phi-psi}b) 
are the product and reactant basins, respectively.
Similar to projection to the ($\phi$-$\theta_1$) plane,  locations of both basins are altered upon projection (SI Table~S2).
Peak 3 becomes very minor after projection to ($\phi-\psi$) plane, and its location is also altered from that of the 5-d surface.  This can be seen in Fig.~\ref{fig:fivetrue-3D}b, where the location of peak 3 on the 5-d surface projected onto ($\phi-\psi$) plane (red point), and the changed location of the new peak after projecting onto the ($\phi-\psi$) plane (blue dot) are shown.

These observations demonstrate that with projection, locations of topological features of probability peaks may change, and their prominence as measured by persistence may also change dramatically. 
Fig.~\ref{fig:fivetrue-3D}c explains why the projection to ($\phi$-$\psi$) results in  the  peak on the 5-d surface (red dot in Fig.~\ref{fig:fivetrue-3D}a) changing location  (Fig.~\ref{fig:fivetrue-3D}b,  peak location shown as a blue dot, red dot no longer at the peak).
First, as shown in Fig.~\ref{fig:fivetrue-3D}c (beginning bar plot), the probability at the location of the red dot
is larger than that at the blue dot on the 5-d surface. 
The probability $p(\phi,\psi)$ at each point on the ($\phi$-$\psi$) plane  (Fig.~\ref{fig:fivetrue-3D}b)  is the sum of all points on the 5-d surface with the same $\phi$ and $\psi$ but with different values in any of the other three coordinates ($\theta_1, \alpha$ and $\beta$). 
Thus, the probability of each point on the ($\phi-\psi$) plane of Fig.~\ref{fig:fivetrue-3D}b is a 3-d hyper-surface  of  $p(\theta_1, \alpha, \beta)$. 
When we  sum up the 3-d hyper-surface  along one direction (\textit{e.g.}, $\alpha$), we obtain a 2-d probability surface (\textit{e.g.} $p(\theta_1, \beta))$.  The 2-d surfaces for the red and blue dots are shown in the middle panel of Fig.~\ref{fig:fivetrue-3D}c.
Reducing the dimension further, we sum up the  2-d probability surface over $\beta$. The resulting distributions are 1-d probability distributions along the $\theta_1$ direction, shown in Fig.~\ref{fig:fivetrue-3D}c for the blue and red dots. 
From these 1-d distributions, we can see that when details in $\theta_1$ direction are retained, the probability at the red dot (actual peak in 5-d) is still higher than the probability at the blue dot. 
We next sum up the probability along the $\theta_1$ direction (Fig.~\ref{fig:fivetrue-3D}c, final bar plots). The  summed values are the probability values over the 2-d ($\phi-\psi$) plane shown in Fig.~\ref{fig:fivetrue-3D}b. 
As shown by the final bar plot in Fig.~\ref{fig:fivetrue-3D}c,   the total probability mass  for the red dot is now less than that for the blue dot, even though  the red dot has a higher probability on the original 5-d surface. 
Hence, the location of peak 3 changes when projecting onto ($\phi-\psi$) plane.

We then carried out committor tests~(Fig.~\ref{fig:margialized_phi-psi}c). None of the topological features are where the transition state ensemble are located:
All trajectories starting from $b_2$, $b_3$, and $d_3$ go to the reactant basin ($p_B = 0)$, 
and all trajectories starting at $b_1$ go to the product basin ($p_B=1.0$).
The committor values for ridge $b_2$ follow a one-sided distribution around $p_B=0$ (Fig.~\ref{fig:margialized_phi-psi}c). Trajectories starting there mostly fall back to the reactant basin. 
Overall, none of the topological features after projection to the ($\phi$-$\psi$) plane  retain the dynamic properties of the transiton state conformations as the original 5-d surface. 

\subsection{Dimension reduction by PCA destroys dynamic properties inherent in surface topology} %
Principal Component Analysis (PCA) is a widely used technique for dimension reduction.  It has found broad applications in  molecular
simulations~\cite{levy1984quasi,garcia1992large,AngularPCA, DihedralPCA1, dihedralPCA2}.
However, whether such reduction retains the essential dynamics of the activation process and whether the surface topology on the PCA  space can uncover the transition state conformations are not known.  Here we assess the dynamic properties of probability surfaces after PCA dimension reduction.

\paragraph{Projection of $p(\phi, \psi, \theta_1,\alpha, \beta,)$ onto $(PC_1, \, PC_2)$ by dPCA.}
We first applied  dihedral 
 principal component analysis (dPCA) 
to the 5-d probability surface~\cite{DihedralPCA1,dihedralPCA2}.
dPCA is widely used for dimension reduction on periodic dimensions.
It first maps each periodic dimension of circular angle to two new dimensions using the $\sin$ and $\cos$ functions. Regular PCA is then applied for dimension reduction.
We use the dPCA procedure and obtain the first two principal components from the variance matrix of  the $1.5\times 10^{11}$ conformations.  Collectively, they account for $80.4\%$ of the variance.

\begin{figure}[h!]

\includegraphics[width=0.99\textwidth]{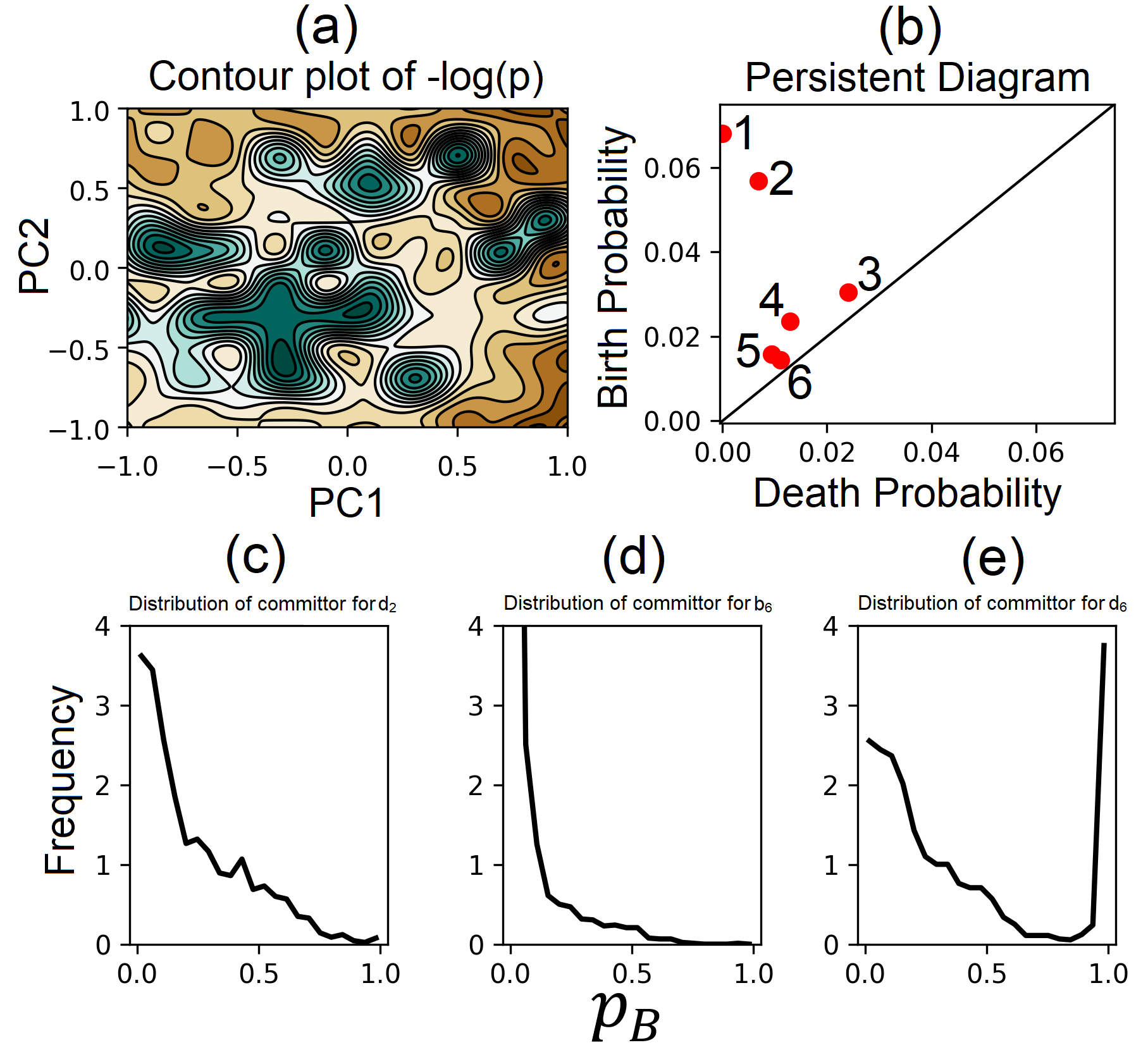}
\caption{\sf The dynamic probability surface on PCA space, its topological structure, and committor values on principal components space. (a) The projection of the 5-d dynamic probability surface $p(\phi, \, \psi, \, \theta_1, \, \alpha, \, \beta)$ to the plane of ($PC_1$-$PC_2$). (b) The persistent diagram exhibits six probability peaks. (c-e) Distribution of committor values $p_B$ for trajectories from  bridge $d_2$, peak $b_6$, and bridge $d_6$, respectively. }
\label{fig:PCA}
\end{figure}

The dynamic probability surface after projection to the ($PC1$-$PC2$) plane is dramatically more complex 
(Fig.~\ref{fig:PCA}a and SI Table~S3) than the surfaces when projected to either the  ($\phi$-$\theta_1$) plane (Fig.~\ref{fig:marginalized}) or the ($\phi$-$\psi$) plane (Fig.~\ref{fig:margialized_phi-psi}). 
The persistent diagram   (Fig.~\ref{fig:PCA}b) is also very different from that of the original 5-d probability surface (Fig.~\ref{fig:fivetrue}b).

The committor tests show that all committor values for conformations at $b_1$ are  1.0 with trajectories going to the product basin. 
All committor values for conformations at $b_{2-5}$ and
$d_{3-5}$ are 0.0: trajectories  from there  all go to the reactant basin. 
Committor values at  ridge $d_2$ and peak $b_6$ follow a one-sided distribution at $p_B=0$ (Fig.~\ref{fig:PCA}c).
Committor values at ridge $d_6$ has higher values both at $p_B=0.0$ and $p_B=1.0$.
Conformations at this location are a mixture of those close to the reactant basin and those close to the product basins.  There are few conformations from the transition state ensemble.

Overall, our results show that the dPCA procedure for dimension reduction removes dynamics relevant information from the topological features of the probability surface. No conformations in the transition state
 of this active process are captured by the topological features of the PCA surface. 

\begin{figure}[h!]

\includegraphics[width=0.99\textwidth]{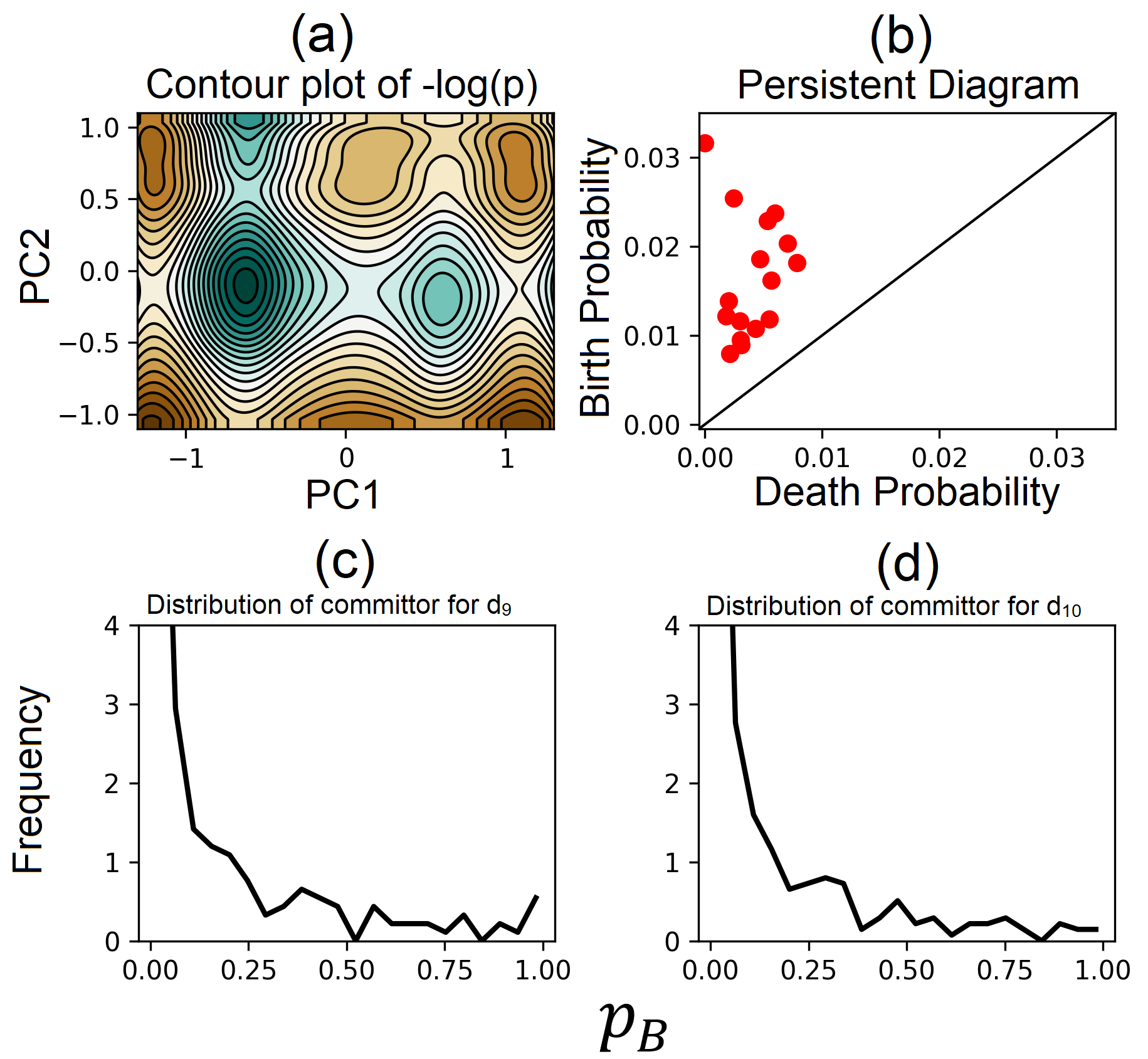}
\caption{\sf The dynamic   probability surface  on 5-d principal components space reduced by dPCA from the 39-d configuration space, its topological structure, and committor values. (a) the 5-d dynamic probability surface $p(PC_1,\cdots,PC_5)$ shown on to the  ($PC_1, PC_2$) plane. (b) The persistent diagram exhibits 16 peaks. (c-d) Distribution of committor  $p_B$ values for trajectories started at ridges $d_9$ and $d_{10}$, respectively. }
\label{fig:PCA2}
\end{figure}

\paragraph{Projection from 39-d angular space to 5-d dPCA subspace.} 
We also applied dPCA to the original full-dimensional dynamic probability surface. After removal of the 21 bond lengths, we apply dPCA to reduce the remaining 39-dimensional  configuration space of angles  to 5 principal components. 
The contour plot of the 5-d PCA  surface  projected onto  the first two principal components is shown in Fig.~\ref{fig:PCA2}a. 
Persistent homology analysis identifies 16 peaks (Fig.~\ref{fig:PCA2}b), each located in a 5-d cube. 
The location and probability of the first 6 most dominant peaks and ridges connecting them are listed in SI Table~S3. 
This 5-d persistent diagram is very different from that shown in Fig.~\ref{fig:fivetrue}b, exhibiting a significantly more complex surface topology.

The committor tests show that all committor values for conformations located at the peaks and the ridges are either 0 or 1.0.
The two exceptions are ridges $d_9$ and $d_{10}$ (Fig.~\ref{fig:PCA2}c-d).
None of the topological features in the dPCA reduced 5-d dynamic probability surface retain the essential dynamics of the transition state ensembles.
Overall, our results demonstrate that 
dimension reduction by dPCA  destroy dynamic properties inherent in the original surface topology
of the dynamic probability surface.

\paragraph{Direct PCA projections.}  Results using direct PCA to project $p(\phi,\, \psi,\, \theta_1,\,\alpha,\,\beta)$ onto ($PC1,\, PC2$), as well as projection from the 39-d space to the 5-d direct PCA subspace  are similar and none retain the dynamic properties in their topological features (see SI for details).

\section {Discussion}
In this study, we have introduced 
a novel approach for characterizing the exact topological features of dynamic probability surfaces.  Instead of 
examining critical points and Morse indexes, ours is based on homology groups of a series of superlevel sets of the probability surface.  With  quantification of the scales of these topological features by persistent homology, we are able to uncover the relationship between the topology of  the dynamic probability surface and the dynamics of the activation process of the alanine-dipeptide isomerization reaction.
 
This approach allows us to define the topological properties of the high-dimensional dynamic probability surface that is associated with the transition state conformations. 
The probability surface over the transition state region
is the most prominent peak after the reactant and product basins.
Instead of a Morse index of $1$ as conventionally thought~\cite{Hanggi1990,morse-1(2)}, transition state ensemble is on the top of a dynamic probability peak and goes downhill in all directions. 
As seen when projected to the ($\phi,\, \theta_1$)-plane (Fig~\ref{fig:fivetrue-3D}a),
it appears as a small peak 
rather than a saddle point commonly associated with transition state on multi-dimensional free energy surface.
This is because the system undergoes certain amount of correlated wandering motions at the barrier top, before it goes down towards the product basin.  Our finding is against the conventional wisdom that the $C_{7eq}\rightarrow C_{7ax}$ transition is a ballistic process, as it is a small peptide and the transition occurs in vacuum.

The dynamic probability surface  was constructed from naturally occurring reactive trajectories  connecting the reactant and product basins.  These trajectories are unbiased and faithfully reflect how the $C_{7eq}\rightarrow C_{7ax}$ transition occurs. 
They contain all the relevant information about the  dynamic process of the activation.  
Unlike the free energy surface commonly used in examining the mechanism of an activated process,
this probability surface 
 contains additional information that reflect the non-equilibrium nature of the transition dynamics.  

\begin{figure}[h!]
    \centering
    \includegraphics[width=0.80\textwidth]{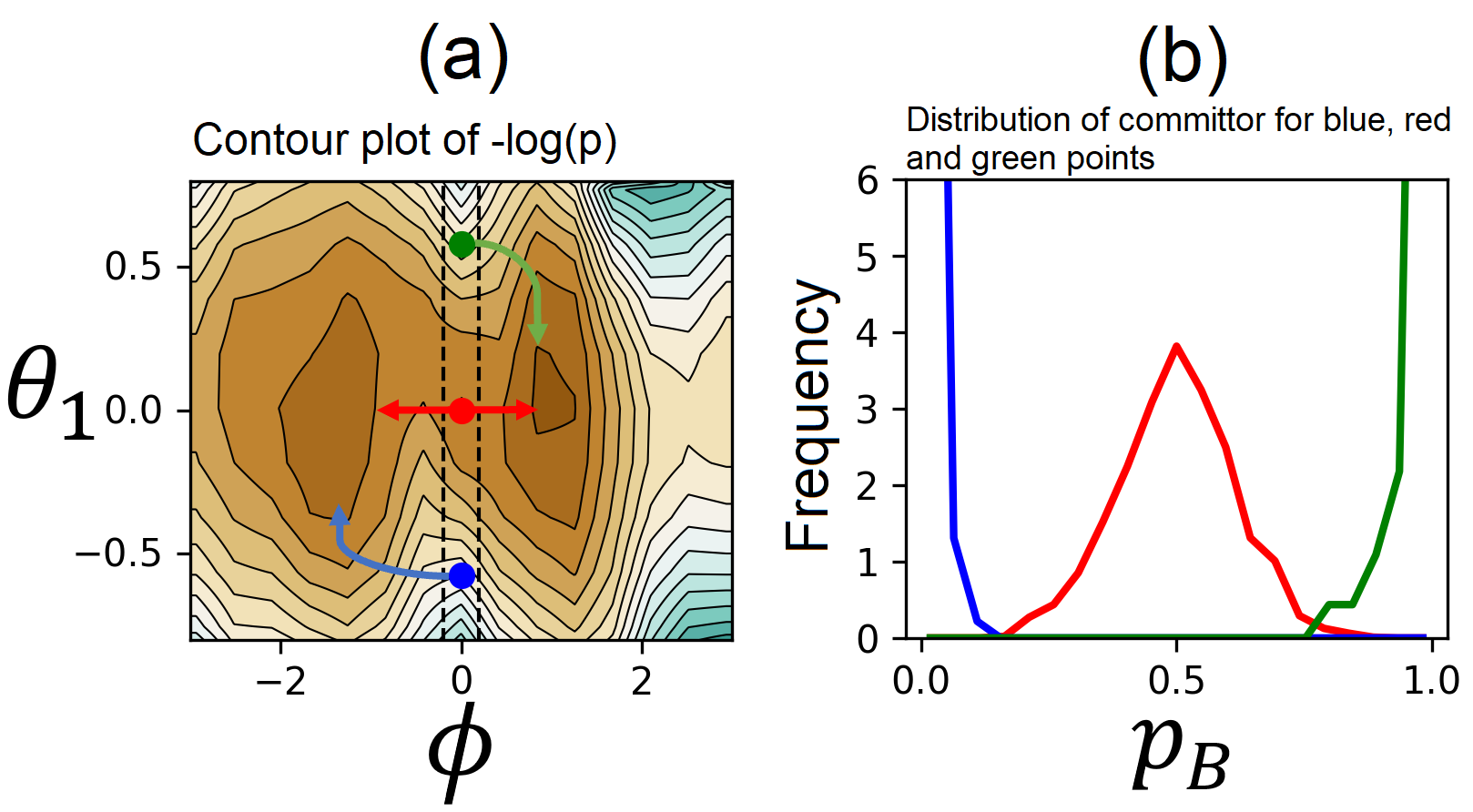}
    \caption{\sf
    The horizontal band of the probability surface of Fig.~\ref{fig:fivenew}a centered at $\theta_2 =0$ ($-0.08<\theta_2<0.08$)  expanded to show distribution  in $\theta_1$. (a) This band containing all peaks in Fig.~\ref{fig:fivenew}a is expanded in the $\theta_1$ direction. 
    The ($\phi$-$\theta_2$)-square containing peak $b_3$ in Fig.~\ref{fig:fivenew}a is expanded in $\theta_1$ and is shown as a  vertical strip  between the two dashed lines. This strip contains a mixture of conformations. The red dot shows the location of the transition state conformations.  (b) The distributions of committor values for conformations at the blue, red, and green dots in (a), respectively. 
    Trajectories from the blue and green dots fall back to the reactant and product basins, respectively.}
    \label{fig:improper_choice}
\end{figure}

A common practice in the studies of protein conformational dynamics is to extract mechanistic insights from the geometry of two-dimensional free energy surface of a
double-well  along certain collective variables, which are often chosen based on  heuristics or by intuition.  For example, one would  associate a saddle region with the transition states.  Our results  illustrate the caveats of such procedures.
All three probability surfaces shown in
Figs.~\ref{fig:fivetrue},~\ref{fig:marginalized}, and~\ref{fig:margialized_phi-psi} exhibit the canonical double-well feature. %
In Fig.~~\ref{fig:marginalized}, the 5-d surface on the ($\phi,\,\theta_2$)-plane  has both the product and reactant peaks and the third peak  laying out along the $\phi$-direction alone, indicating that $\theta_2$ is not a reaction coordinate.  
This is indeed verified by the committor test: configurations corresponding to the peak in the saddle region all share the correct value of $\phi$ but samples randomly along the $\theta_1$ direction,
as illustrated by the flat committor distribution in Fig.~\ref{fig:fivenew}c.  Detialed examination shows that the 5-d cube containing the transition state conformations  
 (red dot in Fig.~\ref{fig:improper_choice}) also contain  conformations with other  $\theta_1$ values, which are not at the transition state: some go to the product basin ({\it e.g.},  green dot, Fig.~\ref{fig:improper_choice}b), and others to the reactant basin ({\it e.g.},  blue dot, Fig.~\ref{fig:improper_choice}b).

In contrast, projection of the 5-d surface on  ($\phi,\,\theta_1$)-plane (Fig~\ref{fig:fivetrue}) has the two basins along the $\phi$-direction alone, but the transition region aligned along both $\phi$ and $\theta_1$.  This second feature is consistent with the importance of $\theta_1$ in determining the barrier crossing
dynamics~\cite{Chandler1978, Kramers1940284, Pechukas1976, Wigner1938, Li_Ma2016_Reaction_Mechanism,Ma2005Automatic,Bolhuis2000}.

Interestingly, the ($\phi,\,\psi$)-surface (Fig~\ref{fig:margialized_phi-psi}) has basins and the saddle region arranged along both directions.  Conventional wisdom would have led to the conclusion that $\psi$ is important in defining both reactant and product basins as well as the barrier crossing process.  However, the double-well structure exhibited on the ($\phi$-$\psi$) plane is profoundly misleading. None of the topological features on the  surface over the ($\phi$-$\psi$) plane  retain the dynamic properties of the original 5-d surface, as the committor test showed that configurations corresponding to the peak at the transition region completely fall into the reactant basin.  This demonstrates that the correlation between $\phi$ and $\psi$, due to minor roles of $\psi$ to the transition
process~\cite{Chandler1978, Kramers1940284}, 
distorted the probability distribution along $\phi$, such that the ridge/saddle extended into the reactant peak, leading to incorrect $\phi$ value that marks the location of the peak in the transition region.  
In contrast, $\theta_2$ did not impact the distribution of $\phi$, so the peak in the transition region on the ($\phi$-$\theta_2$) plane still bears the correct $\phi$ value for the transition states.

In general, the dynamic properties of topological features of the probability surface are very sensitive to the subspace of projection.
This is illustrated by the different locations of the  transition state conformations, which are at a peak location on the ($\phi$-$\theta_1$) plane (Fig~\ref{fig:fivetrue-3D}a,  red dot), but are at
 a slope location  below a new peak when projected to the ($\phi$-$\psi$) plane (Fig~\ref{fig:fivetrue-3D}b, blue dot). 
While the probability  at the correct ($\phi$-$\theta_1$) square for the transition state conformations  (Fig~\ref{fig:fivetrue-3D}c, left, red) is at a peak,  a different location  in the ($\phi$-$\psi$) plane  (Fig.~\ref{fig:fivetrue-3D}b, blue dot) has higher probability, as in this projection the  probability mass distributed along the dimension of $\theta_1$  is integrated over all $\theta_1$ values (Fig.~\ref{fig:fivetrue-3D}c, higher bar, right), resulting in the concentration of probability peak at this new location.

Our results show that without the inclusion of the correct reaction coordinates,
  which can be identified by the energy flow theory~\cite{Li_Ma2016TPS,Li_Ma2016_Reaction_Mechanism}, probability surface of the same dimension no longer correctly characterizes the dynamic properties of the active process.
 While the exact topological features of the  surface will be captured, they no longer correspond to the location of the transition state ensemble.
Without $\theta_1$, the 5-d dynamic probability surface over   
 ($\phi$, $\theta_2$, $\psi$, $\alpha$, $\beta$)  fail to capture the  dynamics of this active process.

Together, our results showed that intuition-based projection (such as $\phi$-$\psi$) or other arbitrary projection cannot be relied upon for understanding the dynamic properties of activated processes.
without  rigorous examination such as the committor test, directly assigning mechanistic significance to features of free energy surface is prone to mistakes, misinterpretations, and misunderstanding.

Finally, our results show that there are dramatic changes in the topological properties of the probability surface after dimension reduction, when techniques such as 
dPCA  are applied.
While the  simple probability surface on
the properly constructed 2-d ($\phi-\theta_1$) plane contains rich dynamic information and is sufficient to uncover the transition state conformations,
the topological features on PCA-reduced surfaces can become more complex and no longer reflect essential dynamics and cannot be used to identify the transition state conformations. 

The approach of homology group and the technique of analyzing the persistent homology of the filtration of the superlevel sets of high-dimensional probability surfaces introduced here 
are general. 
    It also works in principle for systems with more than two reaction coordinates, as long as they are part of the coordinate system, and the cubic complexes can be computed.
We envision they can be applied to investigate  topology of high-dimensional probability surfaces encountered in other physical problems of activated process.

\begin{suppinfo}

Additional information about the birth and death places and probabilities of significant peaks for each situation is available in supporting information. In addition, supporting information provides results from the dimension reduction by direct PCA without using dihedral PCA.

\end{suppinfo}

\begin{acknowledgement}

We thank Drs.\ Herbert Edelsbrunner and Hubert Wagner for discussion and for generous help in extending the cubic complex algorithm.  This work is supported by grants NIH R35 GM127084, NIH R01 GM086536, and NSF CHE-1665104.
\end{acknowledgement}

\section*{Conflict of Interest Statement} 
There are no conflict of interests.

\bibliography{TPS, tda, references, landscapeAnalysis}

\newpage
\setcounter{page}{1}
\setcounter{figure}{0}

\renewcommand{\thepage}{S\arabic{page}} 
\renewcommand{\thetable}{S\arabic{table}}  
\renewcommand{\thefigure}{S\arabic{figure}}

\section{Supporting Information}

\begin{table}[h!]
\fontsize{9}{12}\selectfont
    \centering
    \caption{\sf Locations of birth and death of the probability peaks and the ridges connecting them for dynamic probability surface projected to ($\phi, \psi, \theta_1, \alpha, \beta$) and ($\phi, \psi, \alpha, \beta, \theta_2$) . }
  \begin{tabular}{ | c | l | c || c | l | c |}
    \hline
    \multicolumn{3}{|c||}{$5$-dimensional subspace of ($\phi, \psi, \theta_1, \alpha, \beta$)} &  \multicolumn{3}{c|}{ $5$-dimensional subspace of ($\phi, \psi, \alpha, \beta, \theta_2$) } \\ [1ex]
    \hline \hline
    Label & ($\phi, \psi, \theta_1, \alpha, \beta$) Coordinate & Probability & Label & ($\phi, \psi, \alpha, \beta, \theta_2$) Coordinate & Probability\\ [1ex]
\hline \hline
$b_1$ & ($1.25, -0.84, 0.01, 2.24, 2.00$) & $1.95\times 10^{-3}$ & $b_1$ & ( \, $1.25, \, -0.83\, , \,  2.23 \, , \,  2.00 \, , \, 0.00 \,$) & $1.95\times 10^{-3}$  \\
\hline
$b_2$ & ($ \, -1.68, \,  0.42 \, , \,  -0.18 \, , \,  2.18 \, , \,  1.94 \, $) & $1.00\times 10^{-3}$ & $b_2$ & ($ \, -1.25 \, , \,  0.00 \, , \,  2.18, \,  2.00 \, , \,  0.00 \, $) & $1.19\times 10^{-3}$ \\
\hline
$d_2$ & ($ \, -0.42 \, , \,  -0.42 \, , \,  0.19 \, , \,  2.34 \, , \,  2.15 \, $) & $5.08\times 10^{-4}$ & $d_2$ & ($ \, -0.41 \, , \,  -0.41 \, , \,  2.34 \, , \,  2.15 \, , \,  0.00 \, $) & $4.19\times 10^{-4}$ \\
\hline
$b_3$ & ($ \, 0.00 \, , \,  -0.42 \, , \,  0.00 \, , \,   2.24 \, , \,  2.10 \, $) & $9.98\times 10^{-4}$ & $b_3$ & ($ \, 0 \, , \,  -0.42 \, , \,  2.34 \, , \,  2.15 \, , \,  0.00 \, $) & $8.46\times 10^{-4}$ \\
\hline
$d_3$ & ($ \, 0.42 \, , \,  -0.42 \, , \,  -0.18 \, , \,  2.34 \, , \,  2.15 \, $) & $7.31\times 10^{-4}$ & $d_3$ & ($ \, 0.42 \, , \,  0.42 \, , \,  2.34 \, , \,  2.15 \, , \,  0.00 \, $) & $6.70\times 10^{-4}$ \\
\hline
$b_4$ & ($ \, -2.52 \, , \,  -2.94 \, , \,  0.00 \, , \,  2.18 \, , \,  1.94 \, $) & $1.16\times 10^{-4}$ & $b_4$ & ($ \, -2.52 \, , \,  -2.94 \, , \,  2.18 \, , \,  1.94 \, , \,  0.00 \, $) & $1.02\times 10^{-4}$ \\
\hline
$d_4$ & ($\, -2.52 \, , \,  -1.26 \, ,  \, -0.18 \, , \,   2.18 \, , \,  1.34 \, $) & $6.73\times 10^{-6}$ & $d_4$ & ($ \, -2.51 \, , \,  -1.68 \, , \,  2.18 \, , \,  1.94 \, , \,  0.00 \, $) & $5.82\times 10^{-6}$ \\[1ex] 
\hline 
\end{tabular}
\label{tab:peaks}
\end{table}

\begin{table}[h!]
\fontsize{11}{12}\selectfont
    \centering
    \caption{\sf Locations of birth and death of the probability peaks and the ridges connecting them for the projected 5-d probability landscape to ($\phi-\theta_1$) plane and ($\phi-\psi$) plane. }
  \begin{tabular}{ | c | c | c || c | c | c |}
    \hline
    \multicolumn{3}{|c||}{$p(\phi, \psi, \theta_1, \alpha, \beta)$ projected onto ($\phi-\theta_1$) plane}  &  \multicolumn{3}{c|}{$p(\phi, \psi, \theta_1, \alpha, \beta)$ projected onto ($\phi-\psi$) plane} \\ [1ex]
    \hline \hline
    Label & ($\phi, \theta_1$) Coordinate & Probability & Label & ($\phi, \psi$) Coordinate & Probability\\ [1ex]
\hline \hline
$b_1$ & ($0.84, 0.00$) & $6.17\times 10^{-2}$ & $b_1$ & ( \, $1.25, \, -0.83\, $) & $6.80\times 10^{-2}$  \\
\hline
$b_2$ & ($ \, -1.25, \,  0.00 \,$) & $3.41\times 10^{-2}$ & $b_2$ & ($ \, -1.25 \, , \,  0.00 \, $) & $4.02\times 10^{-2}$ \\
\hline
$d_2$ & ($ \, -0.42 \, 0.19 \,$) & $1.06\times 10^{-2}$ & $d_2$ & ($ \, -0.41 \, , \,  -0.41 \,$) & $8.90\times 10^{-3}$ \\
\hline
$b_3$ & ($ \, 0.00 \, , \,  0.00\, $) & $1.70\times 10^{-2}$ & $b_3$ & ($ \, -2.52 \, , \,  -2.93 \, $) & $3.71\times 10^{-3}$ \\
\hline
$d_3$ & ($ \, 0.42 \, , \,  0.00\, $) & $1.37\times 10^{-4}$ & $d_3$ & ($ \, -2.52 \, , \,  -1.68 \, $) & $1.34\times 10^{-4}$ \\[1ex] 
\hline 
\end{tabular}
\label{tab:peaks_projected}
\end{table}

\begin{table}[h!]
\fontsize{9}{12}\selectfont
    \centering
    \caption{\sf Locations of the probability peaks and ridges connecting them for a) the  5-d probability landscape projected to  2 principal components by dPCA. 
    The two most important eigenvectors $\nu_1$ and $\nu_2$ of the covariance matrix of the conformations are
$\nu_1=[-0.251, \, -0.771, \, -0.112, \, 0.573, \,  -0.003, \, -0.014, \, 0.025, \, 0.019, \, 0.025, 0.011]^T$ and
$\nu_2=[0.339, \,  -0.208, \, 0.910, \, 0.052, -0.003, \, 0.053, \, -0.027, \, -0.022, \, -0.063, \, -0.030]^T$ for ($\cos(\phi),\, \sin(\phi), \, \cos(\psi),\, \sin(\psi), \, \cos(\theta_1),\, \sin(\theta_1), \, \cos(\alpha),\, \sin(\alpha), \, \cos(\beta),\, \sin(\beta)$), with   eigenvalues of $r_1=1.056$ and $r_2=0.283$, respectively. \\
    b) The 39-d probability landscape  projected to the 5 principal components by dPCA.
    $d_9$ shown in text is at $(PC_1,\cdots, PC_5) = (-0.005, \, -0.886, \, 0.958, \, -0.382, \, -0.862)$, and the persistence is  $b_9-d_9 = 1.38\times 10^{-4} - 1.99 \times 10^{-5} = 1.18\times 10^{-4}$. 
$d_{10}$ is  at  $(-0.005, \, 0.882, \, 0.958, \, -0.382, \, -0.862)$ and the persistence is $b_{10}-d_{10}=1.21\times 10^{-4} - 1.80 \times 10^{-5}= 1.03\times 10^{-4}$.
        }
  \begin{tabular}{ | c | c | c || c | c | c |}
    \hline
    \multicolumn{3}{|c||}{$p(\phi, \psi, \theta_1, \alpha, \beta)$ projected onto ($PC_1, PC_2$) plane}  &  \multicolumn{3}{c|}{$p(\bx \in \mathbb{R}^{39})$ projected onto ($PC_i: \, i\in\{1, \cdots, \, 5\}$) plane} \\ [1ex]
    \hline \hline
    Label & ($PC_1, PC_2$) Coordinate & Probability & Label & ($PC_1, \cdots, PC_5$) Coordinate & Probability\\ [1ex]
\hline \hline
$b_1$ & ($-1.3, \, 0.5$) & $6.80\times 10^{-2}$ & $b_1$ & ( \, $-1.35, 0.88, 0.95, -0.38, -0.86$) & $3.16\times 10^{-4}$  \\
\hline
$b_2$ & ($0.7, \,  1.1 \,$) & $5.69\times 10^{-2}$ & $b_2$ & ($ \, -1.35, -0.88, 0.95, -0.38, -0.86 $) & $2.54\times 10^{-4}$ \\
\hline
$d_2$ & ($ \, -0.1 \, 1.3 \,$) & $7.14\times 10^{-3}$ & $d_2$ & ($ \, -1.35, -0.44, 0.95, -0.38, -0.86 \,$) & $2.41\times 10^{-5}$ \\
\hline
$b_3$ & ($ \, 1.1 \, , \,  0.7\, $) & $3.05\times 10^{-2}$ & $b_3$ & ($ \, -1.35, -0.88, -0.38, 0.94, -0.86 \, $) & $2.37\times 10^{-4}$ \\
\hline
$d_3$ & ($ \, 1.1 \, , \,  0.9\, $) & $2.40\times 10^{-2}$ & $d_3$ & ($ \,-1.35, -0.88, 0.95, 0.37, -0.68\, $) & $5.99\times 10^{-5}$ \\
\hline
$b_4$ & ($ \, 1.3 \, , \,  0.3\, $) & $2.36\times 10^{-2}$ & $b_4$ & ($ \,-1.35, 0.88, -0.38, 0.94, -0.86 \, $) & $2.29\times 10^{-4}$ \\
\hline
$d_4$ & ($ \, 1.3 \, , \,  0.5\, $) & $1.29\times 10^{-2}$ & $d_4$ & ($ \, -1.35, 0.88, 1.14, 0.56, -0.51 \, $) & $5.30\times 10^{-5}$ \\
\hline
$b_5$ & ($ \, 1.3 \, , \,  -0.1\, $) & $1.58\times 10^{-2}$ & $b_5$ & ($ \, -1.35, -0.88, -0.95, 0.37, 0.86 \, $) & $2.03\times 10^{-4}$ \\
\hline
$d_5$ & ($ \, 1.3 \, , \,  0.1\, $) & $9.40\times 10^{-3}$ & $d_5$ & ($ \, -1.35, -0.88, 0.00, -1.33, -0.51 \, $) & $7.03\times 10^{-5}$ \\
\hline
$b_6$ & ($ \, -0.3 \, , \,  1.3\, $) & $1.45\times 10^{-2}$ & $b_6$ & ($ \, -1.35, 0.88, 0.38, -0.95, 0.86 \, $) & $1.86\times 10^{-4}$ \\
\hline
$d_6$ & ($ \, -0.5 \, , \,  1.1\, $) & $1.12\times 10^{-2}$ & $d_6$ & ($ \, -1.35, 0.88, 0.19, -1.14, -0.68 \, $) & $4.66\times 10^{-5}$ \\[1ex] 
\hline 
\end{tabular}
\label{tab:pca}
\end{table}

\newpage
\subsection*{Dimension reduction by PCA}

\paragraph{Projection of $p(\phi,\psi,\theta_1,\alpha,\beta)$ onto ($PC1,PC2$) by PCA.} Here we applied direct PCA without using dihedral PCA to the 5-d probability surface. We first shift each direction to its mean value and then perform PCA to obtain the first two principal components from the covariance matrix of the $1.5\times10^{11}$ conformations. 

\begin{figure}[h!]

\includegraphics[width=0.99\textwidth]{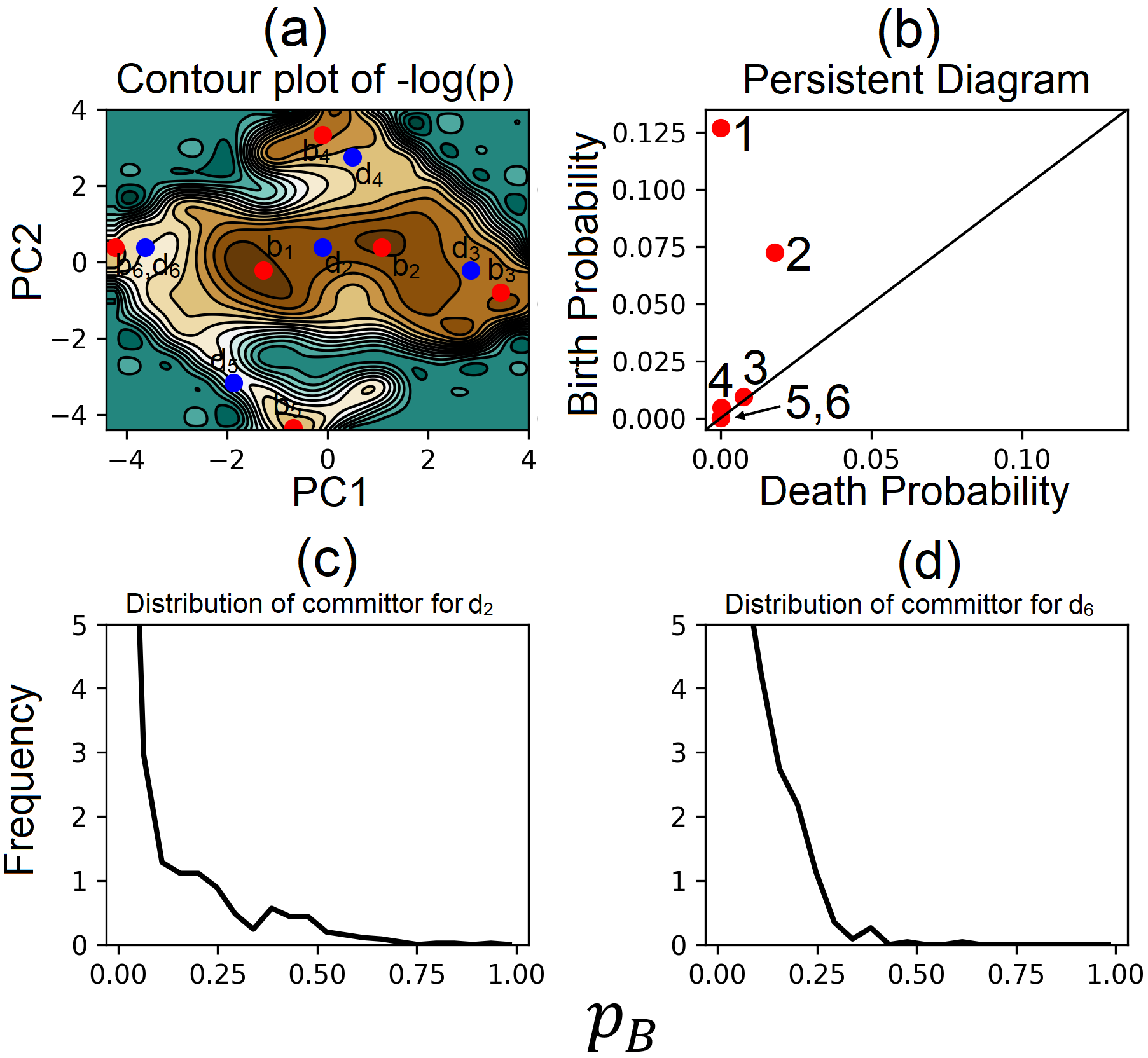}
\caption{\sf The dynamic probability surface on PCA space, its topological structure, and committor values on principal components space. (a) The projection of the 5-d dynamic probability surface $p(\phi, \, \psi, \, \theta_1, \, \alpha, \, \beta)$ to the plane of ($PC_1$-$PC_2$). (b) The persistent diagram exhibits six probability peaks. (c-d) Distribution of committor values $p_B$ for trajectories from  bridge $d_2$, peak $d_6$, respectively. }
\label{fig:AngularPCA}
\end{figure}

The dynamic probability surface after projection to the ($PC1,PC2$) plane is shown in Fig.~\ref{fig:AngularPCA}a. This probability surface is dramatically different than that of the 5-d surface when projected onto ($\phi-\theta_1$) plane or ($\phi-\psi$) plane. The persistence diagram also shows different peaks with different birth and death places (Fig.~\ref{fig:AngularPCA}b). 
Committor tests show that all committor values for the conformations at $b_{2-6}$, and $d_{3-5}$ are 0.0 with trajectories going to the reactant basin. All committor values for conformations at $b_1$ are 1.0: Trajectories from the location of $b_1$ all go to the product basin.
Committor values for ridges $d_2$ and $d_6$ follow a one-sided distribution. Trajectories starting from these locations mostly fall back to the reactant basin. Overall, non of the topological feature after PCA retain dynamic properties of the transition state conformations as the oroginal 5-d surface.

\begin{figure}[h!]

\includegraphics[width=0.99\textwidth]{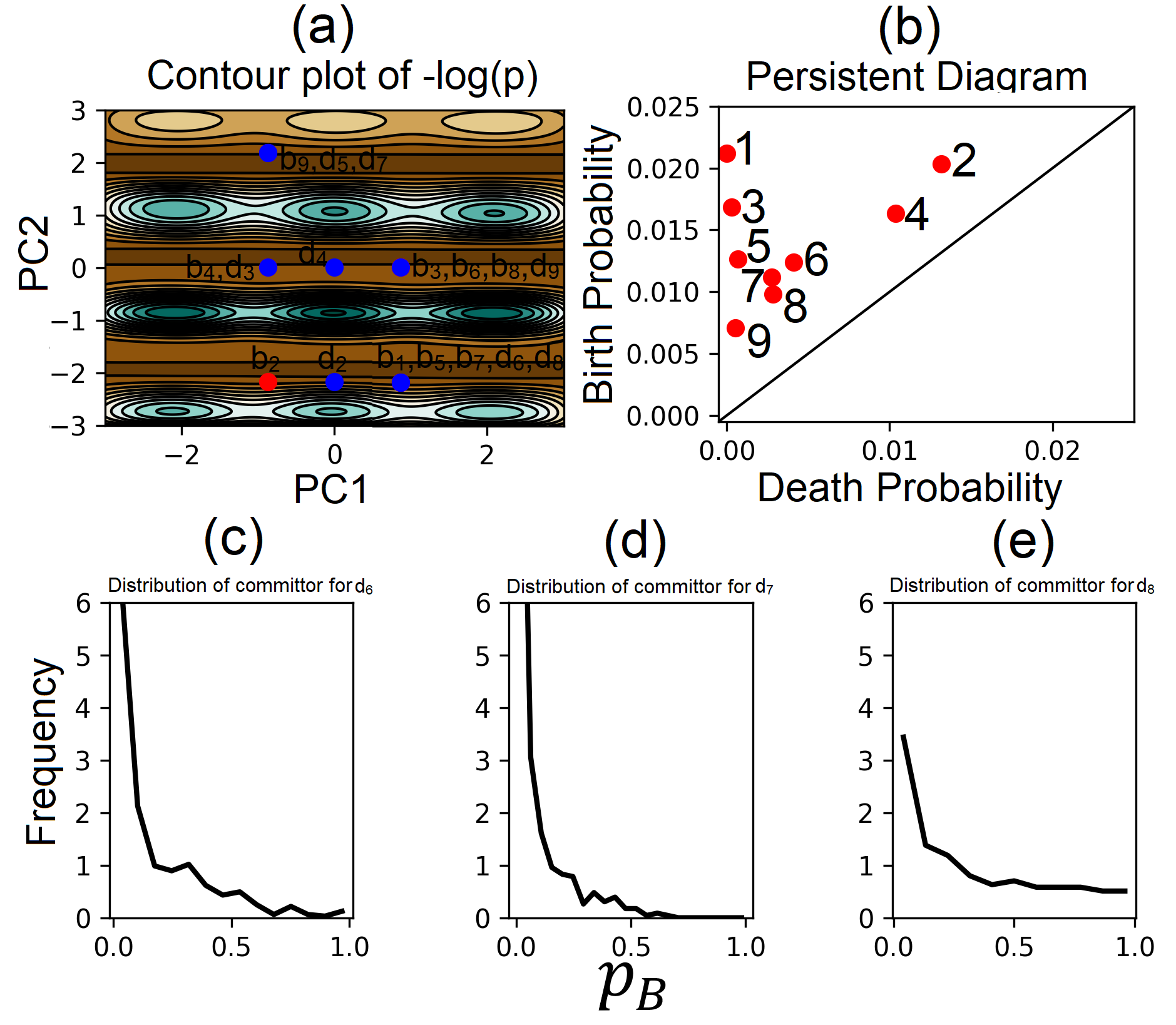}
\caption{\sf The dynamic   probability surface  on 5-d principal components space reduced by dPCA from the 39-d configuration space, its topological structure, and committor values. (a) the 5-d dynamic probability surface $p(PC_1,\cdots,PC_5)$ shown on to the  ($PC_1, PC_2$) plane. (b) The persistent diagram exhibits 9 peaks. (c-e) Distribution of committor  $p_B$ values for trajectories started at ridges $d_6$, $d_{7}$, and $d_8$, respectively. }
\label{fig:AngularPCA39}
\end{figure}

\paragraph{Projecting from 39-d space to 5-d PCA subspace.} We applied direct PCA to the original full dimensional dynamic probability surface, after removal of 21 bond lengths, to reduce the remaining 39-dimensional subspace to 5 principal components. The contour plot of the 5-d PCA surface projected onto the first two principal components is shown in Fig.~\ref{fig:AngularPCA39}a. Persistence homology analysis identifies 9 peaks (Fig.~\ref{fig:AngularPCA39}b), each located in a 5-d cube. These 5-d persistent diagram is very different from the persistence diagram shown for the original 5-d subspace.

The committor tests show that all committor values for conformations located at the peaks $b_{6-8}$ are 0.0. That is all the trajectories starting from these locations fall back to the reactant basin. In addition, all committor values for the conformations located at the peaks $b_{1-5}, b_9$, and the ridges $d_{2-5}, d_9$ are 1.0. Trajectories from these locations move forward to the product basin. 
Committor values for ridges $d_{6-8}$ follow one-sided distributions with the peak at $p_B=0$ (Fig.~\ref{fig:AngularPCA39}c-e). Trajectories starting from these ridges mostly fall back to the reactant basin.

Overall, These results demonstrate that dimension reduction by PCA destroys dynamic properties inherent on the original surface topology of the dynamic probability surface.

\end{document}